\documentclass[  
reprint, 
aps,
amsmath,
amssymb,
amsfonts, 
superscriptaddress,
prx,
longbibliography
]{revtex4-2}

\usepackage{braket}
\usepackage{graphicx}
\usepackage{bm}
\usepackage[colorlinks=true,allcolors=blue]{hyperref} 
\usepackage[T1]{fontenc}

\newcommand{\er}{Er$^{3+}$ }
\newcommand{\upg}{\uparrow}
\newcommand{\downg}{\downarrow}

\newcommand{\panelcol}[1]{\textcolor{blue}{#1}}

\begin{document}

\title{Coherent control of a nuclear spin via interactions with a rare-earth ion in the solid-state}

\author{Mehmet T.~Uysal}
\thanks{These authors contributed equally to this work.}

\author{Mouktik Raha\textsuperscript{*,}}
\altaffiliation{Present address: Pritzker School of Molecular Engineering, The University of Chicago, Chicago, IL 60637, USA}

\author{Songtao Chen}
\altaffiliation{Present address: Department of Electrical and Computer Engineering, Rice University, Houston, TX 77005, USA}

\author{Christopher M.~Phenicie}
\altaffiliation{Present address: Fathom Radiant, Boulder, CO 80301, USA}

\author{Salim Ourari}

\affiliation{\mbox{Department of Electrical and Computer Engineering, Princeton University, Princeton, NJ 08544, USA}}

\author{Mengen Wang}
\author{Chris G.~Van de Walle}

\affiliation{Materials Department, University of California, Santa Barbara, CA 93106-5050, USA}

\author{Viatcheslav V.~Dobrovitski}
\affiliation{QuTech, Delft University of Technology, 2628 CJ Delft, Netherlands}
\affiliation{Kavli Institute of Nanoscience, Delft University of Technology, 2628 CJ Delft, Netherlands}

\author{Jeff D.~Thompson}
\email[]{jdthompson@princeton.edu}
\affiliation{\mbox{Department of Electrical and Computer Engineering, Princeton University, Princeton, NJ 08544, USA}}


\date{\today}

\begin{abstract}
Individually addressed \er ions in solid-state hosts are promising resources for quantum repeaters, because of their direct emission in the telecom band and compatibility with silicon photonic devices. While the \er electron spin provides a spin-photon interface, ancilla nuclear spins could enable multi-qubit registers with longer storage times. In this work, we demonstrate coherent coupling between the electron spin of a single \er ion and a single $I=1/2$ nuclear spin in the solid-state host crystal, which is a fortuitously located proton ($^1$H). We control the nuclear spin using dynamical decoupling sequences applied to the electron spin, implementing one- and two-qubit gate operations. Crucially, the nuclear spin coherence time exceeds the electron coherence time by several orders of magnitude, because of its smaller magnetic moment. These results provide a path towards combining long-lived nuclear spin quantum registers with telecom-wavelength emitters for long-distance quantum repeaters.

\end{abstract}

\maketitle


Optically interfaced solid-state atomic defects are promising for realizing quantum technologies, in particular quantum networks \cite{wehner2018quantum} and sensors \cite{degen2017quantum}. In this approach, the optical transition of the defect provides a direct interface to the electronic spin \cite{awschalom2018quantum}. However, enhanced functionality can be realized by coupling the electronic spin to one or more ancillary nuclear spins in the surrounding environment \cite{jelezko2004observation,childress2006coherent}. For example, in NV centers in diamond, this approach has been used to realize long-lived nuclear spin quantum registers \cite{neumann2008multipartite,maurer2012room}, distribute entangled states across a quantum network \cite{pfaff2014unconditional}, and implement multi-qubit operations such as quantum error correction protocols \cite{taminiau2014universal, waldherr2014quantum, cramer2016repeated, abobeih2022fault}.

\begin{figure*}[!ht]
	\centering
    {\includegraphics[width=172 mm]{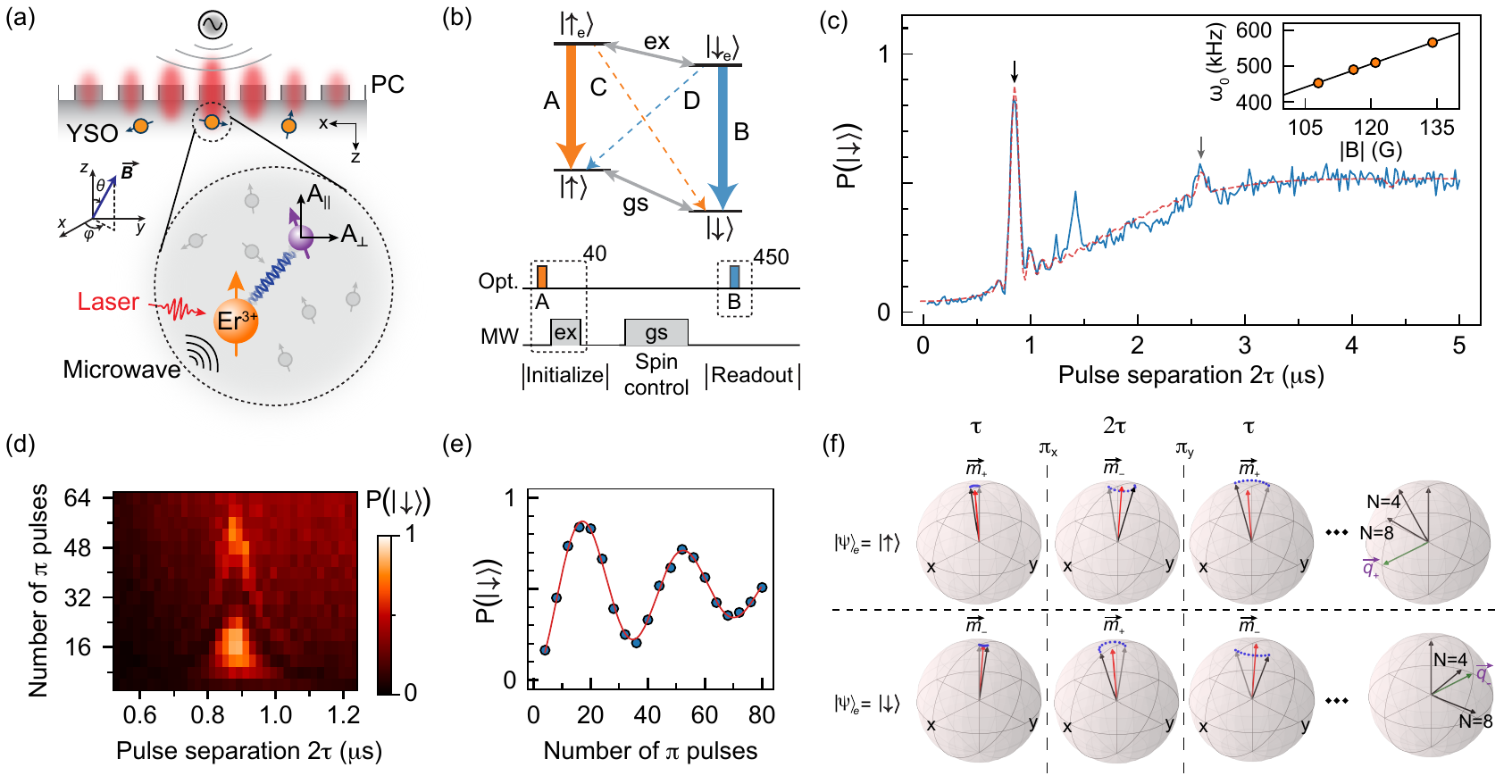}}
    \caption{\label{fig:fig1}\textbf{Detection and coherent control of a weakly-coupled nuclear spin using an \er electronic spin.}
    \textbf{(a)} Measurement scheme. A single \er spin, which is initialized, controlled, and readout via a combination of optical and microwave excitations, probes a nearby nuclear spin.
    \textbf{(b)} (top) \er level structure; (bottom) Pulse scheme for initialization and single-shot readout of \er spin state, interleaved with microwave pulses resonant with the ground state splitting for spin manipulation. Spin-conserving transitions (labeled A,B) are highly cyclic.
    \textbf{(c)} Evolution of an \er spin under an XY-16 sequence as the separation between consecutive $\pi$-pulses ($2\tau$) is varied (solid blue curve). Arrows indicate resonances at $\pi/\omega_0$ and $3\pi/\omega_0$, based on the simulated signal (dashed red line) for the extracted interaction parameters and decay envelope with $T_\text{2,XY16} = 16.1 \pm 0.2 \,\mu$s. Inset: $\omega_0$ as a function of field strength fits to a line with slope $4.3\pm0.1$ kHz/G and offset $-12\pm9$ kHz. Experiments were performed at field angles $(\theta,\phi)$ = $(90^\circ,110^\circ)$ and $(90^\circ,120^\circ)$ in a coordinate system defined by the YSO crystal structure: $(x,y,z) = (D_1,D_2,b)$ \cite{sun2008magnetic}.
    \textbf{(d)} The \er spin population is plotted as a function of $2\tau$ and number of $\pi$-pulses ($N$) in the XY-$N$ sequence near the resonance centered at $2\tau_0 = 0.875\,\mu$s. The ``chevron'' interference pattern originates from coherent interaction between the \er spin and a nuclear spin
    \textbf{(e)} Coherent controlled rotation of the nuclear spin as a function of $N$ at the $2\tau$ resonance position. The data fits to a decaying sinusoidal (solid red line) with $\alpha=10.2^{\circ}$ rotation per $\pi$-pulse.
    \textbf{(f)} Conditional evolution of a nuclear spin based on the \er electronic spin state for one unit ($\tau - \pi_x - 2\tau - \pi_y - \tau$) of our decoupling sequence. 
    Repeating this unit leads to a conditional $x$-rotation of the nuclear spin, around the effective axes $\mathbf{q_{\pm}} = \pm \mathbf{\hat{x}}$.
    All experiments were performed at the magnetic field configuration $(B,\theta,\phi)=(130 \text{ G},95^\circ,110^\circ)$, corresponding to $(f_\text{MW,gs},f_\text{MW,ex}) = (2.02, 1.32)$~GHz, unless indicated otherwise.
    }
\end{figure*}

Individually addressed rare-earth ions (REIs) are an emerging platform \cite{siyushev2014coherent, utikal2014spectroscopic, nakamura2014spectroscopy, dibos2018atomic, zhong2018optically, ulanowski2021spectral} with several attractive features, including demonstrated single-shot spin readout \cite{raha2020optical, kindem2020control}, and sub-wavelength addressing and control of many defects \cite{chen2020parallel}. They can also be incorporated into a range of materials \cite{zhong2019emerging, phenicie2019narrow, stevenson2022erbium}, which allows engineering of their environment to improve coherence and integration with devices \cite{dibos2018atomic,zhong2018optically,yin2013optical}.
Among REIs, \er is particularly promising because its telecom-wavelength (1.5 $\mu$m) optical transition may enable long-distance quantum repeaters without frequency conversion.

REIs are most commonly doped into yttrium-based host materials \cite{liu2006spectroscopic} [\emph{e.g.}, Y$_2$SiO$_5$ (YSO), YVO$_4$ or Y$_3$Al$_5$O$_{12}$], which are magnetically noisy, harboring nuclear spins from Y ($^{89}$Y, with 100\% abundance) and other elements (\emph{e.g.}, $^{29}$Si in YSO with 4.7\% abundance). These form a bath that decoheres the electron spin. Recently, coupling between an REI electron spin (Ce$^{3+}$) and individual nuclear spins in the bath ($^{89}$Y, $^{29}$Si) was observed in Ce:YSO \cite{kornher2020sensing}. Additionally, coherent coupling and quantum gate operations were demonstrated between a single $^{171}$Yb$^{3+}$ electron spin and collective states of a small $^{51}$V nuclear spin ensemble in Yb:YVO$_4$ \cite{ruskuc2022nuclear}. However, the operation of an ancilla nuclear spin qubit with a coherence time longer than the electron spin has not yet been demonstrated in REIs.

In this work, we demonstrate coherent coupling between the electron spin of a single \er ion in YSO to a single nearby nuclear spin. The nuclear spin is controlled via microwave-driven, dynamical decoupling (DD) sequences on the electronic spin, realizing one-qubit and two-qubit gate operations that let us probe the coupling strength and coherent dynamics of the nuclear spin. We observe a nuclear spin $T_2$ time (with a Hahn echo) of 1.9 ms, about three orders of magnitude longer than the same quantity in the \er electron spin alone. With a SWAP operation, we also demonstrate that the state of the nuclear spin in the computational basis survives electron spin readout and re-initialization. The properties of the nuclear spin suggest that it is a fortuitously located proton ($^1$H), whose location we determine by measuring the coupling strength for several magnetic field configurations. This work adds a key component to the toolkit of individually-addressed REIs, and also demonstrates a potentially powerful approach to engineering nuclear spin ancillae using non-native atomic species.


Our experimental platform consists of a silicon nanophotonic cavity (PC) bonded to a single crystal sample of YSO with a trace concentration of \er ions (Fig.~\ref{fig:fig1}\panelcol{a}). The cavity serves to enhance the emission rate \cite{dibos2018atomic} and the cyclicity of the \er optical transitions, enabling single-shot spin readout \cite{raha2020optical}. Additionally, microwave fields from a nearby antenna are used to drive spin rotations with a $\pi$ pulse duration of approximately 31~ns. Additional details about the experimental geometry can be found in Ref.~\cite{chen2021hybrid}.

In a magnetic field, the spin-1/2 ground and excited states of the \er ion become non-degenerate, giving rise to four distinct optical transitions ($A$-$D$) that can be used to control and measure the electron spin (Fig.~\ref{fig:fig1}\panelcol{b}). We initialize the spin by alternately exciting the ion on the $A$ transition and applying microwave $\pi$-pulses to the excited state, which polarizes the spin into $\ket{\downarrow}$ \cite{chen2020parallel}. Spin readout is accomplished by repeatedly exciting the $B$ transition, with a typical single-shot fidelity of $81\%$.

\begin{figure*}[!ht]
	\centering
    \includegraphics[width=172 mm]{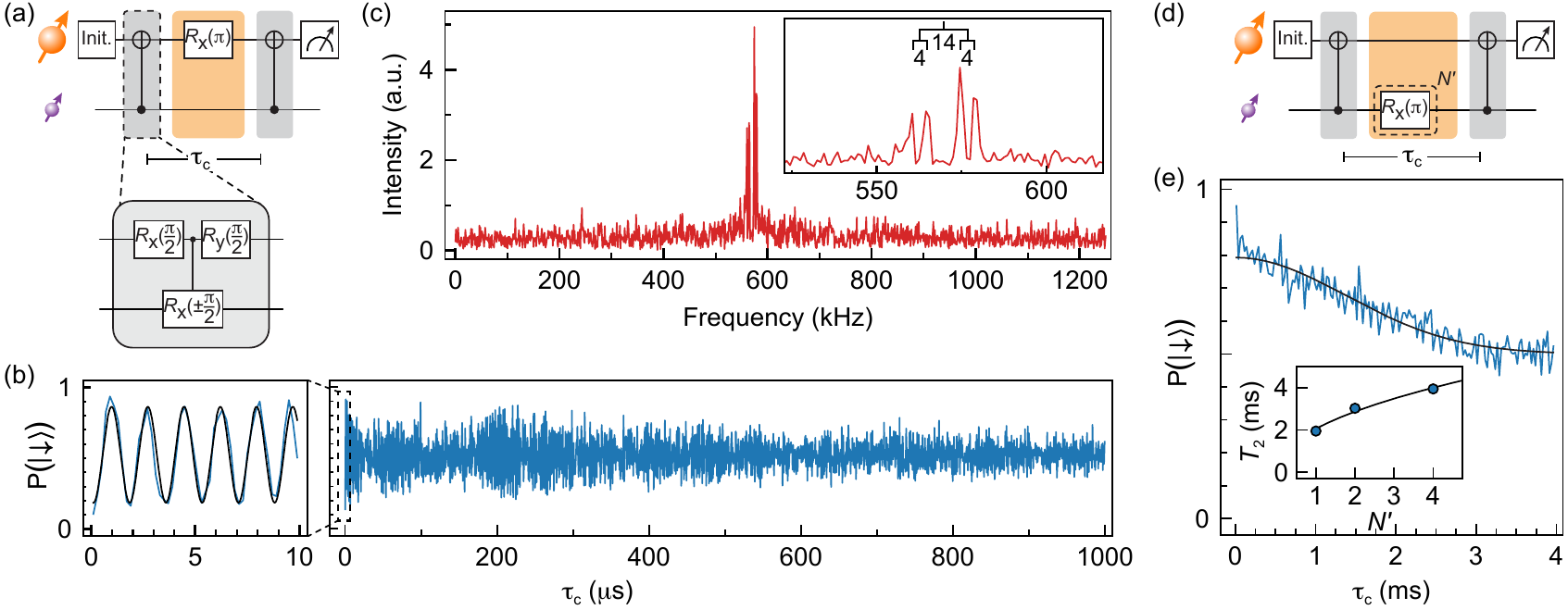}
    \caption{\label{fig:fig2}\textbf{Probing coherence of the nuclear spin.}
    \textbf{(a)} Ramsey spectroscopy of the nuclear spin is performed by changing the delay $\tau_\text{c}$ between two $C_\text{n}NOT_\text{e}$ operations and applying a $\pi$ pulse on the electron at the halfway-point.
    \textbf{(b)} The spectroscopy reveals oscillations at $\omega_0$ in the short time scale (solid black line) and a beating envelope over a millisecond.
    \textbf{(c)} FFT of the Ramsey signal reveals four distinct frequency components that are symmetric around $\omega_0$.
    \textbf{(d)} Hahn and CPMG sequences on the nuclear spin are performed by applying unconditional $R_\text{x}(\pi)$ operations.
    \textbf{(e)} The sequence yields $T_\text{2,Hahn} = 1.9 \pm 0.1$~ms and can be extended to $T_\text{2,CPMG} = 3.9 \pm 0.2$~ms (inset).
    }
\end{figure*}

We probe the local magnetic environment of the \er ion by performing microwave-driven DD sequences $[\text{XY-}N := (\tau-\pi_{x/y}-\tau)^N]$ on the \er spin with varying inter-pulse separation 2$\tau$
\cite{kolkowitz2012sensing,taminiau2012detection,zhao2012sensing}. On top of an overall coherence decay arising from the nuclear spin bath ($T_\text{2,XY16} = 16.1 \pm 0.2 \,\mu$s), the particular \er ion studied in this work exhibits a sharp resonance at $2\tau_0 = 0.875\,\mu$s (Fig.~\ref{fig:fig1}\panelcol{c}). At the resonance, varying the number of $\pi$ pulses $N$ in the XY-$N$ sequence yields coherent oscillations of the \er spin population (Fig.~\ref{fig:fig1}\panelcol{d,e}). We attribute these features to a coherent interaction between the \er electronic spin and a nearby nuclear spin.

To understand this phenomenon, we consider the interaction Hamiltonian between the electronic and nuclear spins in the rotating frame of the electronic spin under the secular approximation:
\begin{equation}
    H/\hbar = 2S_z(A_{||}I_z + A_\perp I_x ) + \omega_\text{L} I_z.
    \label{eq:hyperfine}
\end{equation} 
Here, \textbf{S} is the pseudo spin-1/2 operator for the lowest Kramers' doublet electronic state, \textbf{I} is nuclear spin operator, $A_{||}$ and $A_{\perp}$ are parallel and perpendicular hyperfine coupling parameters, and $\omega_\text{L} = \gamma_\text{N} B/\hbar$ is the (bare) Larmor frequency of the nuclear spin. In the spin Hamiltonian, we define the $\mathbf{\hat{z}}$-axis to be the direction of the external magnetic field, while the $\mathbf{\hat{x}}$ axis is fixed by $A_{\perp}$. The orientation of the field in the reference frame of the YSO crystal axes is specified in Fig.~\ref{fig:fig1}. 
Rearranging terms in Eq.~\eqref{eq:hyperfine}, the Hamiltonian simplifies to:
\begin{equation}
H/\hbar = \ket{\upg}\bra{\upg} \otimes \omega_+\mathbf{I}\cdot\mathbf{m}_+ + \ket{\downg}\bra{\downg} \otimes \omega_-\mathbf{I}\cdot\mathbf{m}_-, 
\label{eq:hyperfine_conditional}
\end{equation}
where $\omega_\pm =$ $\sqrt{(\omega_\text{L} \pm A_{||})^2 + A_{\perp}^2}$ is the effective Larmor frequency and $\mathbf{m}_\pm = (\pm A_\perp, 0, \pm A_{||}+\omega_\text{L})/\omega_{\pm}$ is the precession axis of the nuclear spin. From Eq.~\eqref{eq:hyperfine_conditional}, it is evident that the nuclear spin rotation is conditional on the \er electronic spin state: it precesses around the $\mathbf{m}_+$ ($\mathbf{m}_-$) axis at the  frequency $\omega_+$ ($\omega_-$) when the electron is in the state $\ket{\upg}$ ($\ket{\downg}$).

In the strong magnetic field regime, $\omega_\text{L}$$\gg$$\sqrt{A_{||}^2+A_\perp^2}$, $\mathbf{m}_{\pm}$ are both nearly parallel to the magnetic field axis, $\mathbf{\hat{z}}$. However, the small perpendicular component can be used to generate a conditional rotation of the nuclear spin. Specifically, toggling the electron spin using an XY-$N$ sequence satisfying the resonance condition $2\tau_0=\pi/\omega_0$ [$\omega_0=$ $(\omega_++\omega_-)/2$ $\sim\omega_\text{L}$] leads to a rotation of the nuclear spin around the effective precession axes $\mathbf{q}_{\pm}$, conditional on the initial electronic state (Fig.~\ref{fig:fig1}\panelcol{f}). After $N$ periods, the evolution of the nuclear spin is described by:
\begin{equation}
U_{\text{XY-}N}=\ket{\upg}\bra{\upg} \otimes R_\text{x}(N\alpha)+\ket{\downg}\bra{\downg} \otimes R_\text{x}(-N\alpha),
\label{eq:U_XYN}
\end{equation}
where $\alpha\sim2A_{\perp}/\omega_{\text{L}}$ is the rotation angle per $\pi$ pulse (Fig.~\ref{fig:fig1}\panelcol{e}).
In Fig.~\ref{fig:fig1}\panelcol{c} (inset), we plot the observed resonance frequency of the XY-16 sequence as a function of the magnetic field strength. It exhibits a slope of $4.3\,\pm\,0.1$ kHz/G, which is consistent with the gyromagnetic ratio of a proton (4.258 kHz/G) within the experimental uncertainty. Using measurement techniques described in the Supplemental Material \cite{SI}, we can extract the full set of parameters $(|A_{||}|,A_{\perp}, \omega_\text{L})= (19.4, 50.5, 567.4)$~kHz for the magnetic field orientation used in Fig.~\ref{fig:fig1}\panelcol{c-e}. With these parameters, $\alpha=10.2^{\circ}$, such that XY-8 approximately implements a maximally entangling conditional-$R_\text{x}(\pm \pi/2)$ rotation on the nuclear spin.


We now study the coherence properties of the nuclear spin. Using additional rotations on the electronic spin, the conditional-$R_\text{x}(\pm \pi/2)$ operation is adapted to a $C_\text{n}NOT_\text{e}$ operation, where the nuclear spin acts as a control (Fig.~\ref{fig:fig2}\panelcol{a}). The operation is controlled by the nuclear spin state along the $\mathbf{\hat{x}}$-axis, which is roughly perpendicular to the precession axes $\mathbf{m}_\pm$, which are nearly parallel to the $\mathbf{\hat{z}}$-axis. Therefore, the sequence in Fig.~\ref{fig:fig2}\panelcol{a} measures the precession of the nuclear spin states $\ket{\pm x} = (\ket{\upg}\pm\ket{\downg})/\sqrt{2}$, equivalent to a standard Ramsey measurement of the nuclear spin coherence. The additional $\pi$-pulse on the electronic spin decouples the precession frequency from the (random) initial state of the nuclear spin, so that the average precession frequency $\omega_0$ is observed at short times (Fig.~\ref{fig:fig2}\panelcol{b}).
However, at longer times, the precession shows a complex beating at several frequencies. A fast Fourier transform (FFT) of the Ramsey measurement reveals four symmetrically spaced peaks, where two subgroups of peaks are separated by 14 kHz and each subgroup is split by 4 kHz (Fig.~\ref{fig:fig2}\panelcol{c}). This suggests the presence of additional spins in the nuclear spin environment, which we discuss below.

We measure the nuclear spin $T_2$ with a Hahn echo sequence, by inserting a $\pi$-pulse on the nuclear spin during its precession (Fig.~\ref{fig:fig2}\panelcol{d}). We perform the $\pi$ pulse by applying the conditional-$R_\text{x}(\pm \pi/2)$ twice so that the nuclear spin rotates by $\pi$ around the $\mathbf{\hat{x}}$-axis regardless of the electronic spin state. This cancels the discrete frequency shifts observed in the Ramsey experiment and yields a $T_2$ time of 1.9 ms, which can be extended to 3.9 ms with repeated applications of the nuclear spin $\pi$-pulse in a Carr-Purcell-Meiboom-Gill (CPMG) sequence (Fig.~\ref{fig:fig2}\panelcol{e}). Crucially, this value is three orders of magnitude longer than the bare electronic spin coherence ($T_{2,\text{e}} = 2 \,\mu$s \cite{chen2020parallel}), demonstrating that the smaller nuclear magnetic moment leads to extended coherence even in the magnetically noisy environment of YSO.
\begin{figure}[!t]
	\centering
    \includegraphics[width=86 mm]{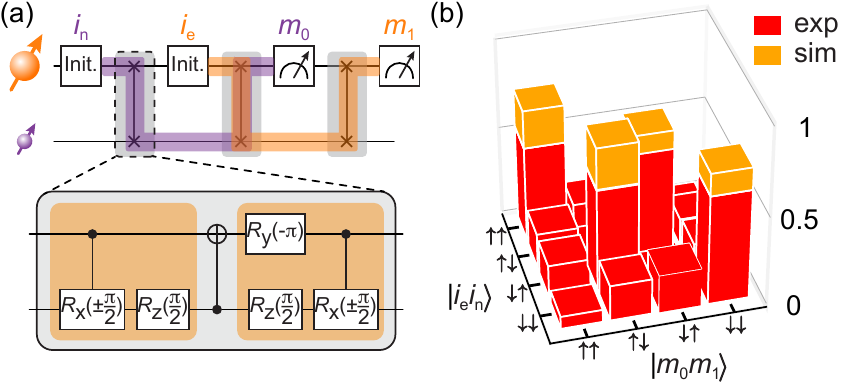}
    \caption{\label{fig:fig3}\textbf{SWAP operation}.
    \textbf{(a)} Demonstration of SWAP operations between the electronic and nuclear spin using the indicated circuit, where the initialization $\ket{i_\text{e}},\ket{i_\text{n}}\in\{\ket{\upg},\ket{\downg}\}$ correspond to nuclear spin and electron spin initialization due to the action of SWAP. This is followed by readout of $\ket{m_0}$ and $\ket{m_1}$ also interleaved with SWAP.
    \textbf{(b)}
    Measurements $\ket{m_0 m_1}$ are plotted for each initialization $\ket{ i_\text{e} i_\text{n}}$. The SWAP pattern is observed as $\ket{i_\text{n}}$ ($\ket{i_\text{e}}$) determines $\ket{m_0}$ ($\ket{m_1}$), after two SWAP operations. The corresponding SWAP fidelity in the computational basis is 87\% (91\%) for the experiment (simulation), where the simulation takes into account known error sources \cite{SI}.
    }
\end{figure}


Next, we implement a SWAP operation to store the electronic state in the nuclear spin. The SWAP consists of conditional-$R_\text{x}(\pm \pi/2)$ rotations interleaved with single qubit rotations of the electronic spin and $\mathbf{\hat{z}}$-rotations of the nuclear spin, realized via its free precession for duration $\tau_0$ (Fig.~\ref{fig:fig3}\panelcol{a}) \cite{SI}. We apply the SWAP operation within a sequence that repeatedly initializes and measures the electronic spin in the $\mathbf{\hat{z}}$-basis. The first initialization $
\ket{i_\text{n}}$ is swapped to the nuclear spin. After completing the second initialization $\ket{i_\text{e}}$, we obtain the two qubit state $\ket{i_\text{e} i_\text{n}}\in\{\ket{\upg\upg},\ket{\upg\downg},\ket{\downg\upg},\ket{\downg\downg}\}$. Performing another SWAP operation before the first readout measures the state of the nuclear spin $\ket{m_0}$, followed by the state of the electronic spin $\ket{m_1}$. The populations measured for each two qubit readout outcome $
\ket{m_0 m_1}$, for a given initialization $\ket{i_\text{e} i_\text{n}}$, yield a SWAP operation fidelity in the computational basis of 87\% (Fig.~\ref{fig:fig3}\panelcol{b}).
This fidelity is consistent with a theoretical estimate incorporating the variation of the nuclear spin precession frequency (Fig.~\ref{fig:fig2}\panelcol{c}), optical excitations of the \er ion and a finite lifetime for the nuclear spin \cite{SI}. This estimate lets us place a lower bound on the intrinsic nuclear spin lifetime of $T_1 > 0.6$ s. However, the changes in the magnetic moment of the electron spin due to excitations during readout generate significant dephasing of the nuclear spin, which precludes storage of arbitrary nuclear spin states during readout in the current configuration.

\begin{figure}[!t]
	\centering
    \includegraphics[width=86 mm]{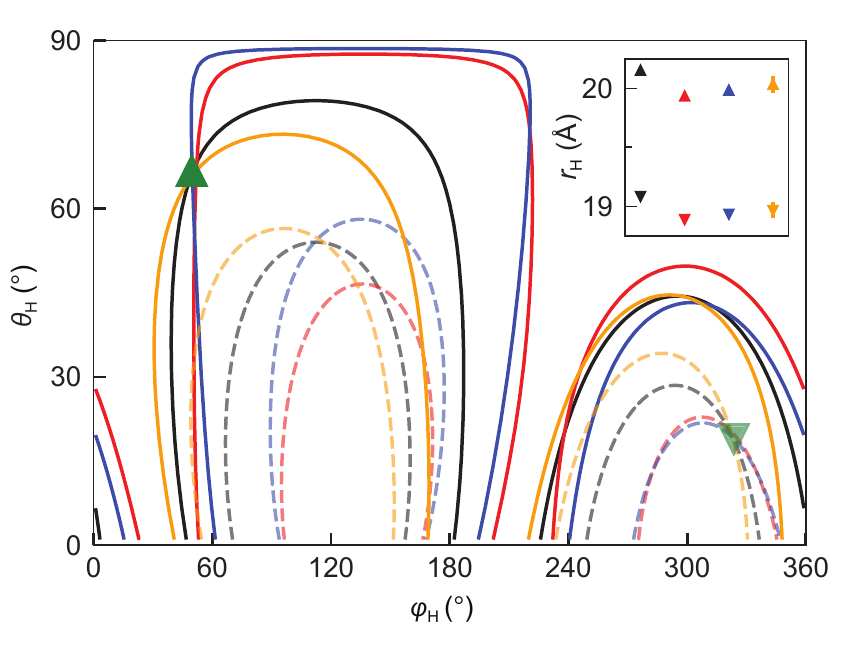}
    \caption{\label{fig:fig4}\textbf{Locating the nuclear spin.}
    $A_{||}/A_{\perp}$ contour plots, where each contour line (black, red, blue, orange) corresponds to nuclear spin positions ($\theta_\text{H}$,$\phi_\text{H}$) that satisfy measurements at magnetic field orientations ($\theta,\phi$) = (95,110), (95,140), (85,140), (100,90) respectively. The intersection of the contour lines, marked by a green triangle, represents a solution. 
    Asserting $A_{||} > 0$ ($A_{||} < 0$), indicated by solid (dashed) contours, leads to a solution at $r_{\text{H}}$,$\theta_\text{H}$,$\phi_\text{H}$ = 20.0 \AA, 66.7$^{\circ}$, 49.6$^{\circ}$ ($r_{\text{H}}$,$\theta_\text{H}$,$\phi_\text{H}$ = 19.0 \AA, 19.0$^{\circ}$, 323.6$^{\circ}$).
    Inset: Calculated distances given $A_{||} > 0$ ($A_{||} < 0$) marked by upward (downward) triangles for each orientation at the indicated solution along with error bars obtained from a Monte-Carlo simulation of measurement uncertainty.
    }
\end{figure}

Finally, we probe the location and gyromagnetic ratio of the nuclear spin in the lattice by measuring the hyperfine parameters  ($A_{||}$,$A_{\perp}$,$\omega_\text{L}$) at four angles of the external magnetic field. The average value of $\omega_{\text{L}}$ yields a gyromagnetic ratio $\gamma_\text{N}/\hbar = 4.26 \pm 0.04$ kHz/G, which is consistent with a $^1$H nuclear spin (4.258 kHz/G). The extracted hyperfine parameters at each field orientation constrain the nuclear spin position to lie on a 1D manifold. The intersection of these curves (Fig.~\ref{fig:fig4}) yields two possible positions of the nuclear spins, which could be further disambiguated by determining the sign of $A_{||}$, which we do not do in this work. For both possible locations, the distance between the electron and nuclear spin is around 1.9 -- 2~nm. We are not able to confidently assign a position in the YSO crystal structure to the nuclear spin, because of uncertainties in the relative alignment of the magnetic axes of the \er site with the crystallographic axes \cite{sun2008magnetic}, which are not aligned because of the low $C_1$ symmetry of the \er site in YSO.

We now turn to a discussion of several aspects of these results. While hydrogen is not part of the YSO chemical formula, it is a ubiquitous impurity that is present during the growth and post-processing and able to unintentionally incorporate in to materials \cite{vandewalle2003H,mccluskey2012hydrogen,shi2004hidden}. The \er center studied in this work is one of six ions that we have probed in the same sample, and the only one to show a strong hydrogen nuclear spin coupling, which is consistent with a hydrogen concentration in the range of 0.3 -- 3.9 $\times 10^{18}~\text{cm}^{-3}$\cite{SI}, a plausible value \cite{devor1984hydroxyl}. Given that hydrogen is quite mobile in oxides, H atoms are also expected to diffuse easily and form defect complexes with large binding energies \cite{mu2022role}. It is also noteworthy that the nuclear spin precession spectrum (Fig.~\ref{fig:fig2}\panelcol{a}) shows four discrete frequency components. In the Supplemental Material, we extend our Ramsey measurements to show that the transitions between these frequencies occur as sudden jumps that can be observed with single-shot measurements, with a correlation time of seconds to minutes \cite{SI}.

We rule out global magnetic field fluctuations and interactions with other nearby electronic spins because these would have a much larger impact on the \er electron, which we do not observe. However, the jumps could be explained by coupling between the H and additional $I=1/2$ nuclei, with Ising interaction strengths of 2 and 7 kHz. In contrast to the distinctively large gyromagnetic ratio of the hydrogen nuclear spin, nuclear spins with smaller gyromagnetic ratios may not be directly observed via the electronic spin within its finite coherence time.

Using comprehensive density functional theory calculations, we conclude that these coupling strengths cannot be explained by the native nuclei alone ($^{89}$Y and $^{29}$Si). However, glow discharge mass spectroscopy (GDMS)  analysis of a related YSO sample reveals the presence of several chemical impurities with $I=1/2$ and ppm concentrations, including Cd and P.
Typically, both interstitial hydrogen (H$_\mathrm{i}$) and substitutional hydrogen (H$_\mathrm{O}$) tend to act as donors in oxides, and are hence attracted to Cd$_\mathrm{Y}$ and P$_\mathrm{O}$, which act as acceptors, which may favor the formation of complexes of these impurities despite their low concentration. This hypothesis is further supported by comprehensive density functional theory calculations: a Cd$_\mathrm{Y}$-H$_\mathrm{O}$ complex and a P$_\mathrm{O}$-H$_\mathrm{i}$ complex, together with a nearby $^{29}$Si, yield coupling strengths to H that are similar to the experimental values (structures and computational details are included in \cite{SI}). Although the observation of such a nuclear spin defect complex is not conclusive at this point, the possibility of engineering nuclear spin registers using this approach is potentially attractive and could be investigated further.


Lastly, we consider the prospects for several extensions to this work. First, driving both the electron and nuclear spins directly would allow the extension of these techniques to multiple nuclear spins \cite{van2012decoherence, bradley2019ten}. Second, to use the nuclear spin as an ancilla in a quantum network, it is necessary to preserve an arbitrary quantum state during a measurement of the electronic spin 
\cite{kalb2018dephasing}. This can be achieved by minimizing the frequency shift of the nuclear spin when the electron is excited to the excited state, either through alignment of the magnetic field, using more weakly coupled nuclear spins, or encoding information in a decoherence-protected subspace of a pair of nuclear spins\cite{reiserer2016robust}.

In conclusion, we demonstrate coherent interaction between a single \er electronic spin and a nuclear spin. Using single and two-qubit operations based on DD sequences, we measure the coherence of this nuclear spin to be several orders of magnitude longer than that of the electronic spin. We further show that the nuclear spin state survives the initialization and readout of the electron in the computational basis by implementing a SWAP operation, and finally, determine its location with respect to the \er ion. Combining the current work with our previous work on parallel measurement and control of multiple \er ions coupled to the same nanophotonic cavity \cite{chen2020parallel}, we can ultimately envision a hybrid quantum register that is comprised of many telecom-compatible optically-interfaced solid-state spins with single particle control at each node, where each spin is individually coupled to long-lived nuclear spin registers.


We acknowledge helpful conversations with Nathalie de Leon. We acknowledge support from AFOSR YIP (FA9550-18-1-0081), DARPA DRINQS (D18AC00015) and a DOE Early Career award (DE-SC0020120, supporting theoretical modeling). V.V.D.~acknowledges support from the DARPA DRINQS program (contract D18AC00015KK1934), as well as the research programme NWO QuTech Physics Funding (QTECH, programme 172) with Project Number 16QTECH02, which is (partly) financed by the Dutch Research Council (NWO) and the Kavli Institute of Nanoscience Delft. For the first-principles calculations, M. W. and C. G. VdW. acknowledge support from the U.S. Department of Energy, Office of Science, National Quantum Information Science Research Centers, Co-design Center for Quantum Advantage (C2QA) under contract number DE-SC0012704; computational resources were provided by the National Energy Research Scientific Computing Center, a DOE Office of Science User Facility supported by the Office of Science of the DOE, under contract No. DE-AC02-05CH11231.


\nocite{pang2005Cgrowth,kresse1996vasp1,kresse1996vasp2,heyd2003hse,oBryan1988yso_lattice}



\bibliography{erbium-nuclei.bib}

\begin{thebibliography}{48}%
\makeatletter
\providecommand \@ifxundefined [1]{%
 \@ifx{#1\undefined}
}%
\providecommand \@ifnum [1]{%
 \ifnum #1\expandafter \@firstoftwo
 \else \expandafter \@secondoftwo
 \fi
}%
\providecommand \@ifx [1]{%
 \ifx #1\expandafter \@firstoftwo
 \else \expandafter \@secondoftwo
 \fi
}%
\providecommand \natexlab [1]{#1}%
\providecommand \enquote  [1]{``#1''}%
\providecommand \bibnamefont  [1]{#1}%
\providecommand \bibfnamefont [1]{#1}%
\providecommand \citenamefont [1]{#1}%
\providecommand \href@noop [0]{\@secondoftwo}%
\providecommand \href [0]{\begingroup \@sanitize@url \@href}%
\providecommand \@href[1]{\@@startlink{#1}\@@href}%
\providecommand \@@href[1]{\endgroup#1\@@endlink}%
\providecommand \@sanitize@url [0]{\catcode `\\12\catcode `\$12\catcode
  `\&12\catcode `\#12\catcode `\^12\catcode `\_12\catcode `\%12\relax}%
\providecommand \@@startlink[1]{}%
\providecommand \@@endlink[0]{}%
\providecommand \url  [0]{\begingroup\@sanitize@url \@url }%
\providecommand \@url [1]{\endgroup\@href {#1}{\urlprefix }}%
\providecommand \urlprefix  [0]{URL }%
\providecommand \Eprint [0]{\href }%
\providecommand \doibase [0]{https://doi.org/}%
\providecommand \selectlanguage [0]{\@gobble}%
\providecommand \bibinfo  [0]{\@secondoftwo}%
\providecommand \bibfield  [0]{\@secondoftwo}%
\providecommand \translation [1]{[#1]}%
\providecommand \BibitemOpen [0]{}%
\providecommand \bibitemStop [0]{}%
\providecommand \bibitemNoStop [0]{.\EOS\space}%
\providecommand \EOS [0]{\spacefactor3000\relax}%
\providecommand \BibitemShut  [1]{\csname bibitem#1\endcsname}%
\let\auto@bib@innerbib\@empty
\bibitem [{\citenamefont {Wehner}\ \emph {et~al.}(2018)\citenamefont {Wehner},
  \citenamefont {Elkouss},\ and\ \citenamefont {Hanson}}]{wehner2018quantum}%
  \BibitemOpen
  \bibfield  {author} {\bibinfo {author} {\bibfnamefont {S.}~\bibnamefont
  {Wehner}}, \bibinfo {author} {\bibfnamefont {D.}~\bibnamefont {Elkouss}},\
  and\ \bibinfo {author} {\bibfnamefont {R.}~\bibnamefont {Hanson}},\
  }\bibfield  {title} {\bibinfo {title} {Quantum internet: A vision for the
  road ahead},\ }\href {https://www.doi.org/10.1126/science.aam9288} {\bibfield
   {journal} {\bibinfo  {journal} {Science}\ }\textbf {\bibinfo {volume}
  {362}},\ \bibinfo {pages} {eaam9288} (\bibinfo {year} {2018})}\BibitemShut
  {NoStop}%
\bibitem [{\citenamefont {Degen}\ \emph {et~al.}(2017)\citenamefont {Degen},
  \citenamefont {Reinhard},\ and\ \citenamefont
  {Cappellaro}}]{degen2017quantum}%
  \BibitemOpen
  \bibfield  {author} {\bibinfo {author} {\bibfnamefont {C.~L.}\ \bibnamefont
  {Degen}}, \bibinfo {author} {\bibfnamefont {F.}~\bibnamefont {Reinhard}},\
  and\ \bibinfo {author} {\bibfnamefont {P.}~\bibnamefont {Cappellaro}},\
  }\bibfield  {title} {\bibinfo {title} {Quantum sensing},\ }\href
  {https://doi.org/10.1103/RevModPhys.89.035002} {\bibfield  {journal}
  {\bibinfo  {journal} {Rev. Mod. Phys.}\ }\textbf {\bibinfo {volume} {89}},\
  \bibinfo {pages} {035002} (\bibinfo {year} {2017})}\BibitemShut {NoStop}%
\bibitem [{\citenamefont {Awschalom}\ \emph {et~al.}(2018)\citenamefont
  {Awschalom}, \citenamefont {Hanson}, \citenamefont {Wrachtrup},\ and\
  \citenamefont {Zhou}}]{awschalom2018quantum}%
  \BibitemOpen
  \bibfield  {author} {\bibinfo {author} {\bibfnamefont {D.~D.}\ \bibnamefont
  {Awschalom}}, \bibinfo {author} {\bibfnamefont {R.}~\bibnamefont {Hanson}},
  \bibinfo {author} {\bibfnamefont {J.}~\bibnamefont {Wrachtrup}},\ and\
  \bibinfo {author} {\bibfnamefont {B.~B.}\ \bibnamefont {Zhou}},\ }\bibfield
  {title} {\bibinfo {title} {Quantum technologies with optically interfaced
  solid-state spins},\ }\href {https://doi.org/10.1038/s41566-018-0232-2}
  {\bibfield  {journal} {\bibinfo  {journal} {Nat. Photonics}\ }\textbf
  {\bibinfo {volume} {12}},\ \bibinfo {pages} {516} (\bibinfo {year}
  {2018})}\BibitemShut {NoStop}%
\bibitem [{\citenamefont {Jelezko}\ \emph {et~al.}(2004)\citenamefont
  {Jelezko}, \citenamefont {Gaebel}, \citenamefont {Popa}, \citenamefont
  {Domhan}, \citenamefont {Gruber},\ and\ \citenamefont
  {Wrachtrup}}]{jelezko2004observation}%
  \BibitemOpen
  \bibfield  {author} {\bibinfo {author} {\bibfnamefont {F.}~\bibnamefont
  {Jelezko}}, \bibinfo {author} {\bibfnamefont {T.}~\bibnamefont {Gaebel}},
  \bibinfo {author} {\bibfnamefont {I.}~\bibnamefont {Popa}}, \bibinfo {author}
  {\bibfnamefont {M.}~\bibnamefont {Domhan}}, \bibinfo {author} {\bibfnamefont
  {A.}~\bibnamefont {Gruber}},\ and\ \bibinfo {author} {\bibfnamefont
  {J.}~\bibnamefont {Wrachtrup}},\ }\bibfield  {title} {\bibinfo {title}
  {Observation of coherent oscillation of a single nuclear spin and realization
  of a two-qubit conditional quantum gate},\ }\href
  {https://doi.org/10.1103/PhysRevLett.93.130501} {\bibfield  {journal}
  {\bibinfo  {journal} {Phys. Rev. Lett.}\ }\textbf {\bibinfo {volume} {93}},\
  \bibinfo {pages} {130501} (\bibinfo {year} {2004})}\BibitemShut {NoStop}%
\bibitem [{\citenamefont {Childress}\ \emph {et~al.}(2006)\citenamefont
  {Childress}, \citenamefont {Gurudev~Dutt}, \citenamefont {Taylor},
  \citenamefont {Zibrov}, \citenamefont {Jelezko}, \citenamefont {Wrachtrup},
  \citenamefont {Hemmer},\ and\ \citenamefont {Lukin}}]{childress2006coherent}%
  \BibitemOpen
  \bibfield  {author} {\bibinfo {author} {\bibfnamefont {L.}~\bibnamefont
  {Childress}}, \bibinfo {author} {\bibfnamefont {M.}~\bibnamefont
  {Gurudev~Dutt}}, \bibinfo {author} {\bibfnamefont {J.}~\bibnamefont
  {Taylor}}, \bibinfo {author} {\bibfnamefont {A.}~\bibnamefont {Zibrov}},
  \bibinfo {author} {\bibfnamefont {F.}~\bibnamefont {Jelezko}}, \bibinfo
  {author} {\bibfnamefont {J.}~\bibnamefont {Wrachtrup}}, \bibinfo {author}
  {\bibfnamefont {P.}~\bibnamefont {Hemmer}},\ and\ \bibinfo {author}
  {\bibfnamefont {M.}~\bibnamefont {Lukin}},\ }\bibfield  {title} {\bibinfo
  {title} {Coherent dynamics of coupled electron and nuclear spin qubits in
  diamond},\ }\href {https://doi.org/10.1126/science.1131871} {\bibfield
  {journal} {\bibinfo  {journal} {Science}\ }\textbf {\bibinfo {volume}
  {314}},\ \bibinfo {pages} {281} (\bibinfo {year} {2006})}\BibitemShut
  {NoStop}%
\bibitem [{\citenamefont {Neumann}\ \emph {et~al.}(2008)\citenamefont
  {Neumann}, \citenamefont {Mizuochi}, \citenamefont {Rempp}, \citenamefont
  {Hemmer}, \citenamefont {Watanabe}, \citenamefont {Yamasaki}, \citenamefont
  {Jacques}, \citenamefont {Gaebel}, \citenamefont {Jelezko},\ and\
  \citenamefont {Wrachtrup}}]{neumann2008multipartite}%
  \BibitemOpen
  \bibfield  {author} {\bibinfo {author} {\bibfnamefont {P.}~\bibnamefont
  {Neumann}}, \bibinfo {author} {\bibfnamefont {N.}~\bibnamefont {Mizuochi}},
  \bibinfo {author} {\bibfnamefont {F.}~\bibnamefont {Rempp}}, \bibinfo
  {author} {\bibfnamefont {P.}~\bibnamefont {Hemmer}}, \bibinfo {author}
  {\bibfnamefont {H.}~\bibnamefont {Watanabe}}, \bibinfo {author}
  {\bibfnamefont {S.}~\bibnamefont {Yamasaki}}, \bibinfo {author}
  {\bibfnamefont {V.}~\bibnamefont {Jacques}}, \bibinfo {author} {\bibfnamefont
  {T.}~\bibnamefont {Gaebel}}, \bibinfo {author} {\bibfnamefont
  {F.}~\bibnamefont {Jelezko}},\ and\ \bibinfo {author} {\bibfnamefont
  {J.}~\bibnamefont {Wrachtrup}},\ }\bibfield  {title} {\bibinfo {title}
  {Multipartite entanglement among single spins in diamond},\ }\href
  {https://doi.org/10.1126/science.1157233} {\bibfield  {journal} {\bibinfo
  {journal} {Science}\ }\textbf {\bibinfo {volume} {320}},\ \bibinfo {pages}
  {1326} (\bibinfo {year} {2008})}\BibitemShut {NoStop}%
\bibitem [{\citenamefont {Maurer}\ \emph {et~al.}(2012)\citenamefont {Maurer},
  \citenamefont {Kucsko}, \citenamefont {Latta}, \citenamefont {Jiang},
  \citenamefont {Yao}, \citenamefont {Bennett}, \citenamefont {Pastawski},
  \citenamefont {Hunger}, \citenamefont {Chisholm}, \citenamefont {Markham}
  \emph {et~al.}}]{maurer2012room}%
  \BibitemOpen
  \bibfield  {author} {\bibinfo {author} {\bibfnamefont {P.~C.}\ \bibnamefont
  {Maurer}}, \bibinfo {author} {\bibfnamefont {G.}~\bibnamefont {Kucsko}},
  \bibinfo {author} {\bibfnamefont {C.}~\bibnamefont {Latta}}, \bibinfo
  {author} {\bibfnamefont {L.}~\bibnamefont {Jiang}}, \bibinfo {author}
  {\bibfnamefont {N.~Y.}\ \bibnamefont {Yao}}, \bibinfo {author} {\bibfnamefont
  {S.~D.}\ \bibnamefont {Bennett}}, \bibinfo {author} {\bibfnamefont
  {F.}~\bibnamefont {Pastawski}}, \bibinfo {author} {\bibfnamefont
  {D.}~\bibnamefont {Hunger}}, \bibinfo {author} {\bibfnamefont
  {N.}~\bibnamefont {Chisholm}}, \bibinfo {author} {\bibfnamefont
  {M.}~\bibnamefont {Markham}}, \emph {et~al.},\ }\bibfield  {title} {\bibinfo
  {title} {Room-temperature quantum bit memory exceeding one second},\ }\href
  {https://doi.org/10.1126/science.1220513} {\bibfield  {journal} {\bibinfo
  {journal} {Science}\ }\textbf {\bibinfo {volume} {336}},\ \bibinfo {pages}
  {1283} (\bibinfo {year} {2012})}\BibitemShut {NoStop}%
\bibitem [{\citenamefont {Pfaff}\ \emph {et~al.}(2014)\citenamefont {Pfaff},
  \citenamefont {Hensen}, \citenamefont {Bernien}, \citenamefont {van Dam},
  \citenamefont {Blok}, \citenamefont {Taminiau}, \citenamefont {Tiggelman},
  \citenamefont {Schouten}, \citenamefont {Markham}, \citenamefont {Twitchen},\
  and\ \citenamefont {Hanson}}]{pfaff2014unconditional}%
  \BibitemOpen
  \bibfield  {author} {\bibinfo {author} {\bibfnamefont {W.}~\bibnamefont
  {Pfaff}}, \bibinfo {author} {\bibfnamefont {B.~J.}\ \bibnamefont {Hensen}},
  \bibinfo {author} {\bibfnamefont {H.}~\bibnamefont {Bernien}}, \bibinfo
  {author} {\bibfnamefont {S.~B.}\ \bibnamefont {van Dam}}, \bibinfo {author}
  {\bibfnamefont {M.~S.}\ \bibnamefont {Blok}}, \bibinfo {author}
  {\bibfnamefont {T.~H.}\ \bibnamefont {Taminiau}}, \bibinfo {author}
  {\bibfnamefont {M.~J.}\ \bibnamefont {Tiggelman}}, \bibinfo {author}
  {\bibfnamefont {R.~N.}\ \bibnamefont {Schouten}}, \bibinfo {author}
  {\bibfnamefont {M.}~\bibnamefont {Markham}}, \bibinfo {author} {\bibfnamefont
  {D.~J.}\ \bibnamefont {Twitchen}},\ and\ \bibinfo {author} {\bibfnamefont
  {R.}~\bibnamefont {Hanson}},\ }\bibfield  {title} {\bibinfo {title}
  {Unconditional quantum teleportation between distant solid-state quantum
  bits},\ }\href {https://doi.org/10.1126/science.1253512} {\bibfield
  {journal} {\bibinfo  {journal} {Science}\ }\textbf {\bibinfo {volume}
  {345}},\ \bibinfo {pages} {532} (\bibinfo {year} {2014})}\BibitemShut
  {NoStop}%
\bibitem [{\citenamefont {Taminiau}\ \emph {et~al.}(2014)\citenamefont
  {Taminiau}, \citenamefont {Cramer}, \citenamefont {van~der Sar},
  \citenamefont {Dobrovitski},\ and\ \citenamefont
  {Hanson}}]{taminiau2014universal}%
  \BibitemOpen
  \bibfield  {author} {\bibinfo {author} {\bibfnamefont {T.~H.}\ \bibnamefont
  {Taminiau}}, \bibinfo {author} {\bibfnamefont {J.}~\bibnamefont {Cramer}},
  \bibinfo {author} {\bibfnamefont {T.}~\bibnamefont {van~der Sar}}, \bibinfo
  {author} {\bibfnamefont {V.~V.}\ \bibnamefont {Dobrovitski}},\ and\ \bibinfo
  {author} {\bibfnamefont {R.}~\bibnamefont {Hanson}},\ }\bibfield  {title}
  {\bibinfo {title} {Universal control and error correction in multi-qubit spin
  registers in diamond},\ }\href {https://doi.org/10.1038/nnano.2014.2}
  {\bibfield  {journal} {\bibinfo  {journal} {Nat. Nanotechnol.}\ }\textbf
  {\bibinfo {volume} {9}},\ \bibinfo {pages} {171} (\bibinfo {year}
  {2014})}\BibitemShut {NoStop}%
\bibitem [{\citenamefont {Waldherr}\ \emph {et~al.}(2014)\citenamefont
  {Waldherr}, \citenamefont {Wang}, \citenamefont {Zaiser}, \citenamefont
  {Jamali}, \citenamefont {Schulte-Herbr{\"u}ggen}, \citenamefont {Abe},
  \citenamefont {Ohshima}, \citenamefont {Isoya}, \citenamefont {Du},
  \citenamefont {Neumann} \emph {et~al.}}]{waldherr2014quantum}%
  \BibitemOpen
  \bibfield  {author} {\bibinfo {author} {\bibfnamefont {G.}~\bibnamefont
  {Waldherr}}, \bibinfo {author} {\bibfnamefont {Y.}~\bibnamefont {Wang}},
  \bibinfo {author} {\bibfnamefont {S.}~\bibnamefont {Zaiser}}, \bibinfo
  {author} {\bibfnamefont {M.}~\bibnamefont {Jamali}}, \bibinfo {author}
  {\bibfnamefont {T.}~\bibnamefont {Schulte-Herbr{\"u}ggen}}, \bibinfo {author}
  {\bibfnamefont {H.}~\bibnamefont {Abe}}, \bibinfo {author} {\bibfnamefont
  {T.}~\bibnamefont {Ohshima}}, \bibinfo {author} {\bibfnamefont
  {J.}~\bibnamefont {Isoya}}, \bibinfo {author} {\bibfnamefont
  {J.}~\bibnamefont {Du}}, \bibinfo {author} {\bibfnamefont {P.}~\bibnamefont
  {Neumann}}, \emph {et~al.},\ }\bibfield  {title} {\bibinfo {title} {Quantum
  error correction in a solid-state hybrid spin register},\ }\href
  {https://doi.org/10.1038/nature12919} {\bibfield  {journal} {\bibinfo
  {journal} {Nature}\ }\textbf {\bibinfo {volume} {506}},\ \bibinfo {pages}
  {204} (\bibinfo {year} {2014})}\BibitemShut {NoStop}%
\bibitem [{\citenamefont {Cramer}\ \emph {et~al.}(2016)\citenamefont {Cramer},
  \citenamefont {Kalb}, \citenamefont {Rol}, \citenamefont {Hensen},
  \citenamefont {Blok}, \citenamefont {Markham}, \citenamefont {Twitchen},
  \citenamefont {Hanson},\ and\ \citenamefont {Taminiau}}]{cramer2016repeated}%
  \BibitemOpen
  \bibfield  {author} {\bibinfo {author} {\bibfnamefont {J.}~\bibnamefont
  {Cramer}}, \bibinfo {author} {\bibfnamefont {N.}~\bibnamefont {Kalb}},
  \bibinfo {author} {\bibfnamefont {M.~A.}\ \bibnamefont {Rol}}, \bibinfo
  {author} {\bibfnamefont {B.}~\bibnamefont {Hensen}}, \bibinfo {author}
  {\bibfnamefont {M.~S.}\ \bibnamefont {Blok}}, \bibinfo {author}
  {\bibfnamefont {M.}~\bibnamefont {Markham}}, \bibinfo {author} {\bibfnamefont
  {D.~J.}\ \bibnamefont {Twitchen}}, \bibinfo {author} {\bibfnamefont
  {R.}~\bibnamefont {Hanson}},\ and\ \bibinfo {author} {\bibfnamefont {T.~H.}\
  \bibnamefont {Taminiau}},\ }\bibfield  {title} {\bibinfo {title} {Repeated
  quantum error correction on a continuously encoded qubit by real-time
  feedback},\ }\href {https://doi.org/10.1038/ncomms11526} {\bibfield
  {journal} {\bibinfo  {journal} {Nat. Commun.}\ }\textbf {\bibinfo {volume}
  {7}},\ \bibinfo {pages} {11526} (\bibinfo {year} {2016})}\BibitemShut
  {NoStop}%
\bibitem [{\citenamefont {Abobeih}\ \emph {et~al.}(2022)\citenamefont
  {Abobeih}, \citenamefont {Wang}, \citenamefont {Randall}, \citenamefont
  {Loenen}, \citenamefont {Bradley}, \citenamefont {Markham}, \citenamefont
  {Twitchen}, \citenamefont {Terhal},\ and\ \citenamefont
  {Taminiau}}]{abobeih2022fault}%
  \BibitemOpen
  \bibfield  {author} {\bibinfo {author} {\bibfnamefont {M.~H.}\ \bibnamefont
  {Abobeih}}, \bibinfo {author} {\bibfnamefont {Y.}~\bibnamefont {Wang}},
  \bibinfo {author} {\bibfnamefont {J.}~\bibnamefont {Randall}}, \bibinfo
  {author} {\bibfnamefont {S.~J.~H.}\ \bibnamefont {Loenen}}, \bibinfo {author}
  {\bibfnamefont {C.~E.}\ \bibnamefont {Bradley}}, \bibinfo {author}
  {\bibfnamefont {M.}~\bibnamefont {Markham}}, \bibinfo {author} {\bibfnamefont
  {D.~J.}\ \bibnamefont {Twitchen}}, \bibinfo {author} {\bibfnamefont {B.~M.}\
  \bibnamefont {Terhal}},\ and\ \bibinfo {author} {\bibfnamefont {T.~H.}\
  \bibnamefont {Taminiau}},\ }\bibfield  {title} {\bibinfo {title}
  {Fault-tolerant operation of a logical qubit in a diamond quantum
  processor},\ }\href {https://doi.org/10.1038/s41586-022-04819-6} {\bibfield
  {journal} {\bibinfo  {journal} {Nature}\ }\textbf {\bibinfo {volume} {606}},\
  \bibinfo {pages} {884} (\bibinfo {year} {2022})}\BibitemShut {NoStop}%
\bibitem [{\citenamefont {Sun}\ \emph {et~al.}(2008)\citenamefont {Sun},
  \citenamefont {B{\"o}ttger}, \citenamefont {Thiel},\ and\ \citenamefont
  {Cone}}]{sun2008magnetic}%
  \BibitemOpen
  \bibfield  {author} {\bibinfo {author} {\bibfnamefont {Y.}~\bibnamefont
  {Sun}}, \bibinfo {author} {\bibfnamefont {T.}~\bibnamefont {B{\"o}ttger}},
  \bibinfo {author} {\bibfnamefont {C.}~\bibnamefont {Thiel}},\ and\ \bibinfo
  {author} {\bibfnamefont {R.}~\bibnamefont {Cone}},\ }\bibfield  {title}
  {\bibinfo {title} {Magnetic $g$ tensors for the $^4\text{I}_{15/2}$ and
  $^4\text{I}_{13/2}$ states of
  $\text{Er}^{3+}\!\!:\!\!\text{Y}_2\text{SiO}_5$},\ }\href
  {https://doi.org/10.1103/PhysRevB.77.085124} {\bibfield  {journal} {\bibinfo
  {journal} {Phys. Rev. B}\ }\textbf {\bibinfo {volume} {77}},\ \bibinfo
  {pages} {085124} (\bibinfo {year} {2008})}\BibitemShut {NoStop}%
\bibitem [{\citenamefont {Siyushev}\ \emph {et~al.}(2014)\citenamefont
  {Siyushev}, \citenamefont {Xia}, \citenamefont {Reuter}, \citenamefont
  {Jamali}, \citenamefont {Zhao}, \citenamefont {Yang}, \citenamefont {Duan},
  \citenamefont {Kukharchyk}, \citenamefont {Wieck}, \citenamefont {Kolesov}
  \emph {et~al.}}]{siyushev2014coherent}%
  \BibitemOpen
  \bibfield  {author} {\bibinfo {author} {\bibfnamefont {P.}~\bibnamefont
  {Siyushev}}, \bibinfo {author} {\bibfnamefont {K.}~\bibnamefont {Xia}},
  \bibinfo {author} {\bibfnamefont {R.}~\bibnamefont {Reuter}}, \bibinfo
  {author} {\bibfnamefont {M.}~\bibnamefont {Jamali}}, \bibinfo {author}
  {\bibfnamefont {N.}~\bibnamefont {Zhao}}, \bibinfo {author} {\bibfnamefont
  {N.}~\bibnamefont {Yang}}, \bibinfo {author} {\bibfnamefont {C.}~\bibnamefont
  {Duan}}, \bibinfo {author} {\bibfnamefont {N.}~\bibnamefont {Kukharchyk}},
  \bibinfo {author} {\bibfnamefont {A.}~\bibnamefont {Wieck}}, \bibinfo
  {author} {\bibfnamefont {R.}~\bibnamefont {Kolesov}}, \emph {et~al.},\
  }\bibfield  {title} {\bibinfo {title} {Coherent properties of single
  rare-earth spin qubits},\ }\href {https://doi.org/10.1038/ncomms4895}
  {\bibfield  {journal} {\bibinfo  {journal} {Nat. Commun.}\ }\textbf {\bibinfo
  {volume} {5}},\ \bibinfo {pages} {3895} (\bibinfo {year} {2014})}\BibitemShut
  {NoStop}%
\bibitem [{\citenamefont {Utikal}\ \emph {et~al.}(2014)\citenamefont {Utikal},
  \citenamefont {Eichhammer}, \citenamefont {Petersen}, \citenamefont {Renn},
  \citenamefont {G{\"o}tzinger},\ and\ \citenamefont
  {Sandoghdar}}]{utikal2014spectroscopic}%
  \BibitemOpen
  \bibfield  {author} {\bibinfo {author} {\bibfnamefont {T.}~\bibnamefont
  {Utikal}}, \bibinfo {author} {\bibfnamefont {E.}~\bibnamefont {Eichhammer}},
  \bibinfo {author} {\bibfnamefont {L.}~\bibnamefont {Petersen}}, \bibinfo
  {author} {\bibfnamefont {A.}~\bibnamefont {Renn}}, \bibinfo {author}
  {\bibfnamefont {S.}~\bibnamefont {G{\"o}tzinger}},\ and\ \bibinfo {author}
  {\bibfnamefont {V.}~\bibnamefont {Sandoghdar}},\ }\bibfield  {title}
  {\bibinfo {title} {Spectroscopic detection and state preparation of a single
  praseodymium ion in a crystal},\ }\href {https://doi.org/10.1038/ncomms4627}
  {\bibfield  {journal} {\bibinfo  {journal} {Nat. Commun.}\ }\textbf {\bibinfo
  {volume} {5}},\ \bibinfo {pages} {3627} (\bibinfo {year} {2014})}\BibitemShut
  {NoStop}%
\bibitem [{\citenamefont {Nakamura}\ \emph {et~al.}(2014)\citenamefont
  {Nakamura}, \citenamefont {Yoshihiro}, \citenamefont {Inagawa}, \citenamefont
  {Fujiyoshi},\ and\ \citenamefont {Matsushita}}]{nakamura2014spectroscopy}%
  \BibitemOpen
  \bibfield  {author} {\bibinfo {author} {\bibfnamefont {I.}~\bibnamefont
  {Nakamura}}, \bibinfo {author} {\bibfnamefont {T.}~\bibnamefont {Yoshihiro}},
  \bibinfo {author} {\bibfnamefont {H.}~\bibnamefont {Inagawa}}, \bibinfo
  {author} {\bibfnamefont {S.}~\bibnamefont {Fujiyoshi}},\ and\ \bibinfo
  {author} {\bibfnamefont {M.}~\bibnamefont {Matsushita}},\ }\bibfield  {title}
  {\bibinfo {title} {Spectroscopy of single $\text{Pr}^{3+}$ ion in
  $\text{LaF}_3$ crystal at 1.5 $\text{K}$},\ }\href
  {https://doi.org/10.1038/srep07364} {\bibfield  {journal} {\bibinfo
  {journal} {Sci. Rep.}\ }\textbf {\bibinfo {volume} {4}},\ \bibinfo {pages}
  {7364} (\bibinfo {year} {2014})}\BibitemShut {NoStop}%
\bibitem [{\citenamefont {Dibos}\ \emph {et~al.}(2018)\citenamefont {Dibos},
  \citenamefont {Raha}, \citenamefont {Phenicie},\ and\ \citenamefont
  {Thompson}}]{dibos2018atomic}%
  \BibitemOpen
  \bibfield  {author} {\bibinfo {author} {\bibfnamefont {A.}~\bibnamefont
  {Dibos}}, \bibinfo {author} {\bibfnamefont {M.}~\bibnamefont {Raha}},
  \bibinfo {author} {\bibfnamefont {C.}~\bibnamefont {Phenicie}},\ and\
  \bibinfo {author} {\bibfnamefont {J.~D.}\ \bibnamefont {Thompson}},\
  }\bibfield  {title} {\bibinfo {title} {Atomic source of single photons in the
  telecom band},\ }\href {https://doi.org/10.1103/PhysRevLett.120.243601}
  {\bibfield  {journal} {\bibinfo  {journal} {Phys. Rev. Lett.}\ }\textbf
  {\bibinfo {volume} {120}},\ \bibinfo {pages} {243601} (\bibinfo {year}
  {2018})}\BibitemShut {NoStop}%
\bibitem [{\citenamefont {Zhong}\ \emph {et~al.}(2018)\citenamefont {Zhong},
  \citenamefont {Kindem}, \citenamefont {Bartholomew}, \citenamefont {Rochman},
  \citenamefont {Craiciu}, \citenamefont {Verma}, \citenamefont {Nam},
  \citenamefont {Marsili}, \citenamefont {Shaw}, \citenamefont {Beyer} \emph
  {et~al.}}]{zhong2018optically}%
  \BibitemOpen
  \bibfield  {author} {\bibinfo {author} {\bibfnamefont {T.}~\bibnamefont
  {Zhong}}, \bibinfo {author} {\bibfnamefont {J.~M.}\ \bibnamefont {Kindem}},
  \bibinfo {author} {\bibfnamefont {J.~G.}\ \bibnamefont {Bartholomew}},
  \bibinfo {author} {\bibfnamefont {J.}~\bibnamefont {Rochman}}, \bibinfo
  {author} {\bibfnamefont {I.}~\bibnamefont {Craiciu}}, \bibinfo {author}
  {\bibfnamefont {V.}~\bibnamefont {Verma}}, \bibinfo {author} {\bibfnamefont
  {S.~W.}\ \bibnamefont {Nam}}, \bibinfo {author} {\bibfnamefont
  {F.}~\bibnamefont {Marsili}}, \bibinfo {author} {\bibfnamefont {M.~D.}\
  \bibnamefont {Shaw}}, \bibinfo {author} {\bibfnamefont {A.~D.}\ \bibnamefont
  {Beyer}}, \emph {et~al.},\ }\bibfield  {title} {\bibinfo {title} {Optically
  addressing single rare-earth ions in a nanophotonic cavity},\ }\href
  {https://doi.org/10.1103/PhysRevLett.121.183603} {\bibfield  {journal}
  {\bibinfo  {journal} {Phys. Rev. Lett.}\ }\textbf {\bibinfo {volume} {121}},\
  \bibinfo {pages} {183603} (\bibinfo {year} {2018})}\BibitemShut {NoStop}%
\bibitem [{\citenamefont {Ulanowski}\ \emph {et~al.}(2021)\citenamefont
  {Ulanowski}, \citenamefont {Merkel},\ and\ \citenamefont
  {Reiserer}}]{ulanowski2021spectral}%
  \BibitemOpen
  \bibfield  {author} {\bibinfo {author} {\bibfnamefont {A.}~\bibnamefont
  {Ulanowski}}, \bibinfo {author} {\bibfnamefont {B.}~\bibnamefont {Merkel}},\
  and\ \bibinfo {author} {\bibfnamefont {A.}~\bibnamefont {Reiserer}},\
  }\bibfield  {title} {\bibinfo {title} {Spectral multiplexing of telecom
  emitters with stable transition frequency},\ }\href
  {https://doi.org/10.48550/arXiv.2110.09409} {\bibfield  {journal} {\bibinfo
  {journal} {arXiv:2110.09409 \!\!\!\!}\ } (\bibinfo {year}
  {2021})}\BibitemShut {NoStop}%
\bibitem [{\citenamefont {Raha}\ \emph {et~al.}(2020)\citenamefont {Raha},
  \citenamefont {Chen}, \citenamefont {Phenicie}, \citenamefont {Ourari},
  \citenamefont {Dibos},\ and\ \citenamefont {Thompson}}]{raha2020optical}%
  \BibitemOpen
  \bibfield  {author} {\bibinfo {author} {\bibfnamefont {M.}~\bibnamefont
  {Raha}}, \bibinfo {author} {\bibfnamefont {S.}~\bibnamefont {Chen}}, \bibinfo
  {author} {\bibfnamefont {C.~M.}\ \bibnamefont {Phenicie}}, \bibinfo {author}
  {\bibfnamefont {S.}~\bibnamefont {Ourari}}, \bibinfo {author} {\bibfnamefont
  {A.~M.}\ \bibnamefont {Dibos}},\ and\ \bibinfo {author} {\bibfnamefont
  {J.~D.}\ \bibnamefont {Thompson}},\ }\bibfield  {title} {\bibinfo {title}
  {Optical quantum nondemolition measurement of a single rare earth ion
  qubit},\ }\href {https://doi.org/10.1038/s41467-020-15138-7} {\bibfield
  {journal} {\bibinfo  {journal} {Nat. Commun.}\ }\textbf {\bibinfo {volume}
  {11}},\ \bibinfo {pages} {1605} (\bibinfo {year} {2020})}\BibitemShut
  {NoStop}%
\bibitem [{\citenamefont {Kindem}\ \emph {et~al.}(2020)\citenamefont {Kindem},
  \citenamefont {Ruskuc}, \citenamefont {Bartholomew}, \citenamefont {Rochman},
  \citenamefont {Huan},\ and\ \citenamefont {Faraon}}]{kindem2020control}%
  \BibitemOpen
  \bibfield  {author} {\bibinfo {author} {\bibfnamefont {J.~M.}\ \bibnamefont
  {Kindem}}, \bibinfo {author} {\bibfnamefont {A.}~\bibnamefont {Ruskuc}},
  \bibinfo {author} {\bibfnamefont {J.~G.}\ \bibnamefont {Bartholomew}},
  \bibinfo {author} {\bibfnamefont {J.}~\bibnamefont {Rochman}}, \bibinfo
  {author} {\bibfnamefont {Y.~Q.}\ \bibnamefont {Huan}},\ and\ \bibinfo
  {author} {\bibfnamefont {A.}~\bibnamefont {Faraon}},\ }\bibfield  {title}
  {\bibinfo {title} {Control and single-shot readout of an ion embedded in a
  nanophotonic cavity},\ }\href {https://doi.org/10.1038/s41586-020-2160-9}
  {\bibfield  {journal} {\bibinfo  {journal} {Nature}\ }\textbf {\bibinfo
  {volume} {580}},\ \bibinfo {pages} {201} (\bibinfo {year}
  {2020})}\BibitemShut {NoStop}%
\bibitem [{\citenamefont {Chen}\ \emph {et~al.}(2020)\citenamefont {Chen},
  \citenamefont {Raha}, \citenamefont {Phenicie}, \citenamefont {Ourari},\ and\
  \citenamefont {Thompson}}]{chen2020parallel}%
  \BibitemOpen
  \bibfield  {author} {\bibinfo {author} {\bibfnamefont {S.}~\bibnamefont
  {Chen}}, \bibinfo {author} {\bibfnamefont {M.}~\bibnamefont {Raha}}, \bibinfo
  {author} {\bibfnamefont {C.~M.}\ \bibnamefont {Phenicie}}, \bibinfo {author}
  {\bibfnamefont {S.}~\bibnamefont {Ourari}},\ and\ \bibinfo {author}
  {\bibfnamefont {J.~D.}\ \bibnamefont {Thompson}},\ }\bibfield  {title}
  {\bibinfo {title} {Parallel single-shot measurement and coherent control of
  solid-state spins below the diffraction limit},\ }\href
  {https://doi.org/10.1126/science.abc7821} {\bibfield  {journal} {\bibinfo
  {journal} {Science}\ }\textbf {\bibinfo {volume} {370}},\ \bibinfo {pages}
  {592} (\bibinfo {year} {2020})}\BibitemShut {NoStop}%
\bibitem [{\citenamefont {Zhong}\ and\ \citenamefont
  {Goldner}(2019)}]{zhong2019emerging}%
  \BibitemOpen
  \bibfield  {author} {\bibinfo {author} {\bibfnamefont {T.}~\bibnamefont
  {Zhong}}\ and\ \bibinfo {author} {\bibfnamefont {P.}~\bibnamefont
  {Goldner}},\ }\bibfield  {title} {\bibinfo {title} {Emerging rare-earth doped
  material platforms for quantum nanophotonics},\ }\href
  {https://doi.org/10.1515/nanoph-2019-0185} {\bibfield  {journal} {\bibinfo
  {journal} {Nanophotonics}\ }\textbf {\bibinfo {volume} {8}},\ \bibinfo
  {pages} {2003} (\bibinfo {year} {2019})}\BibitemShut {NoStop}%
\bibitem [{\citenamefont {Phenicie}\ \emph {et~al.}(2019)\citenamefont
  {Phenicie}, \citenamefont {Stevenson}, \citenamefont {Welinski},
  \citenamefont {Rose}, \citenamefont {Asfaw}, \citenamefont {Cava},
  \citenamefont {Lyon}, \citenamefont {De~Leon},\ and\ \citenamefont
  {Thompson}}]{phenicie2019narrow}%
  \BibitemOpen
  \bibfield  {author} {\bibinfo {author} {\bibfnamefont {C.~M.}\ \bibnamefont
  {Phenicie}}, \bibinfo {author} {\bibfnamefont {P.}~\bibnamefont {Stevenson}},
  \bibinfo {author} {\bibfnamefont {S.}~\bibnamefont {Welinski}}, \bibinfo
  {author} {\bibfnamefont {B.~C.}\ \bibnamefont {Rose}}, \bibinfo {author}
  {\bibfnamefont {A.~T.}\ \bibnamefont {Asfaw}}, \bibinfo {author}
  {\bibfnamefont {R.~J.}\ \bibnamefont {Cava}}, \bibinfo {author}
  {\bibfnamefont {S.~A.}\ \bibnamefont {Lyon}}, \bibinfo {author}
  {\bibfnamefont {N.~P.}\ \bibnamefont {De~Leon}},\ and\ \bibinfo {author}
  {\bibfnamefont {J.~D.}\ \bibnamefont {Thompson}},\ }\bibfield  {title}
  {\bibinfo {title} {Narrow optical line widths in erbium implanted in
  $\text{TiO}_2$},\ }\href {https://doi.org/10.1021/acs.nanolett.9b03831}
  {\bibfield  {journal} {\bibinfo  {journal} {Nano Lett.}\ }\textbf {\bibinfo
  {volume} {19}},\ \bibinfo {pages} {8928} (\bibinfo {year}
  {2019})}\BibitemShut {NoStop}%
\bibitem [{\citenamefont {Stevenson}\ \emph {et~al.}(2022)\citenamefont
  {Stevenson}, \citenamefont {Phenicie}, \citenamefont {Gray}, \citenamefont
  {Horvath}, \citenamefont {Welinski}, \citenamefont {Ferrenti}, \citenamefont
  {Ferrier}, \citenamefont {Goldner}, \citenamefont {Das}, \citenamefont
  {Ramesh} \emph {et~al.}}]{stevenson2022erbium}%
  \BibitemOpen
  \bibfield  {author} {\bibinfo {author} {\bibfnamefont {P.}~\bibnamefont
  {Stevenson}}, \bibinfo {author} {\bibfnamefont {C.~M.}\ \bibnamefont
  {Phenicie}}, \bibinfo {author} {\bibfnamefont {I.}~\bibnamefont {Gray}},
  \bibinfo {author} {\bibfnamefont {S.~P.}\ \bibnamefont {Horvath}}, \bibinfo
  {author} {\bibfnamefont {S.}~\bibnamefont {Welinski}}, \bibinfo {author}
  {\bibfnamefont {A.~M.}\ \bibnamefont {Ferrenti}}, \bibinfo {author}
  {\bibfnamefont {A.}~\bibnamefont {Ferrier}}, \bibinfo {author} {\bibfnamefont
  {P.}~\bibnamefont {Goldner}}, \bibinfo {author} {\bibfnamefont
  {S.}~\bibnamefont {Das}}, \bibinfo {author} {\bibfnamefont {R.}~\bibnamefont
  {Ramesh}}, \emph {et~al.},\ }\bibfield  {title} {\bibinfo {title}
  {Erbium-implanted materials for quantum communication applications},\ }\href
  {https://doi.org/10.1103/PhysRevB.105.224106} {\bibfield  {journal} {\bibinfo
   {journal} {Phys. Rev. B}\ }\textbf {\bibinfo {volume} {105}},\ \bibinfo
  {pages} {224106} (\bibinfo {year} {2022})}\BibitemShut {NoStop}%
\bibitem [{\citenamefont {Yin}\ \emph {et~al.}(2013)\citenamefont {Yin},
  \citenamefont {Rancic}, \citenamefont {de~Boo}, \citenamefont {Stavrias},
  \citenamefont {McCallum}, \citenamefont {Sellars},\ and\ \citenamefont
  {Rogge}}]{yin2013optical}%
  \BibitemOpen
  \bibfield  {author} {\bibinfo {author} {\bibfnamefont {C.}~\bibnamefont
  {Yin}}, \bibinfo {author} {\bibfnamefont {M.}~\bibnamefont {Rancic}},
  \bibinfo {author} {\bibfnamefont {G.~G.}\ \bibnamefont {de~Boo}}, \bibinfo
  {author} {\bibfnamefont {N.}~\bibnamefont {Stavrias}}, \bibinfo {author}
  {\bibfnamefont {J.~C.}\ \bibnamefont {McCallum}}, \bibinfo {author}
  {\bibfnamefont {M.~J.}\ \bibnamefont {Sellars}},\ and\ \bibinfo {author}
  {\bibfnamefont {S.}~\bibnamefont {Rogge}},\ }\bibfield  {title} {\bibinfo
  {title} {Optical addressing of an individual erbium ion in silicon},\ }\href
  {https://doi.org/10.1038/nature12081} {\bibfield  {journal} {\bibinfo
  {journal} {Nature}\ }\textbf {\bibinfo {volume} {497}},\ \bibinfo {pages}
  {91} (\bibinfo {year} {2013})}\BibitemShut {NoStop}%
\bibitem [{\citenamefont {Liu}\ and\ \citenamefont
  {Jacquier}(2005)}]{liu2006spectroscopic}%
  \BibitemOpen
  \bibfield  {author} {\bibinfo {author} {\bibfnamefont {G.}~\bibnamefont
  {Liu}}\ and\ \bibinfo {author} {\bibfnamefont {B.}~\bibnamefont {Jacquier}},\
  }\href {https://doi.org/10.1007/3-540-28209-2} {\emph {\bibinfo {title}
  {Spectroscopic $\text{P}$roperties of $\text{R}$are $\text{E}$arths in
  $\text{O}$ptical $\text{M}$aterials}}}\ (\bibinfo  {publisher} {Springer
  Berlin, Heidelberg},\ \bibinfo {year} {2005})\BibitemShut {NoStop}%
\bibitem [{\citenamefont {Kornher}\ \emph {et~al.}(2020)\citenamefont
  {Kornher}, \citenamefont {Xiao}, \citenamefont {Xia}, \citenamefont {Sardi},
  \citenamefont {Zhao}, \citenamefont {Kolesov},\ and\ \citenamefont
  {Wrachtrup}}]{kornher2020sensing}%
  \BibitemOpen
  \bibfield  {author} {\bibinfo {author} {\bibfnamefont {T.}~\bibnamefont
  {Kornher}}, \bibinfo {author} {\bibfnamefont {D.-W.}\ \bibnamefont {Xiao}},
  \bibinfo {author} {\bibfnamefont {K.}~\bibnamefont {Xia}}, \bibinfo {author}
  {\bibfnamefont {F.}~\bibnamefont {Sardi}}, \bibinfo {author} {\bibfnamefont
  {N.}~\bibnamefont {Zhao}}, \bibinfo {author} {\bibfnamefont {R.}~\bibnamefont
  {Kolesov}},\ and\ \bibinfo {author} {\bibfnamefont {J.}~\bibnamefont
  {Wrachtrup}},\ }\bibfield  {title} {\bibinfo {title} {Sensing individual
  nuclear spins with a single rare-earth electron spin},\ }\href
  {https://doi.org/10.1103/PhysRevLett.124.170402} {\bibfield  {journal}
  {\bibinfo  {journal} {Phys. Rev. Lett.}\ }\textbf {\bibinfo {volume} {124}},\
  \bibinfo {pages} {170402} (\bibinfo {year} {2020})}\BibitemShut {NoStop}%
\bibitem [{\citenamefont {Ruskuc}\ \emph {et~al.}(2022)\citenamefont {Ruskuc},
  \citenamefont {Wu}, \citenamefont {Rochman}, \citenamefont {Choi},\ and\
  \citenamefont {Faraon}}]{ruskuc2022nuclear}%
  \BibitemOpen
  \bibfield  {author} {\bibinfo {author} {\bibfnamefont {A.}~\bibnamefont
  {Ruskuc}}, \bibinfo {author} {\bibfnamefont {C.-J.}\ \bibnamefont {Wu}},
  \bibinfo {author} {\bibfnamefont {J.}~\bibnamefont {Rochman}}, \bibinfo
  {author} {\bibfnamefont {J.}~\bibnamefont {Choi}},\ and\ \bibinfo {author}
  {\bibfnamefont {A.}~\bibnamefont {Faraon}},\ }\bibfield  {title} {\bibinfo
  {title} {Nuclear spin-wave quantum register for a solid-state qubit},\ }\href
  {https://doi.org/10.1038/s41586-021-04293-6} {\bibfield  {journal} {\bibinfo
  {journal} {Nature}\ }\textbf {\bibinfo {volume} {602}},\ \bibinfo {pages}
  {408} (\bibinfo {year} {2022})}\BibitemShut {NoStop}%
\bibitem [{\citenamefont {Chen}\ \emph {et~al.}(2021)\citenamefont {Chen},
  \citenamefont {Ourari}, \citenamefont {Raha}, \citenamefont {Phenicie},
  \citenamefont {Uysal},\ and\ \citenamefont {Thompson}}]{chen2021hybrid}%
  \BibitemOpen
  \bibfield  {author} {\bibinfo {author} {\bibfnamefont {S.}~\bibnamefont
  {Chen}}, \bibinfo {author} {\bibfnamefont {S.}~\bibnamefont {Ourari}},
  \bibinfo {author} {\bibfnamefont {M.}~\bibnamefont {Raha}}, \bibinfo {author}
  {\bibfnamefont {C.~M.}\ \bibnamefont {Phenicie}}, \bibinfo {author}
  {\bibfnamefont {M.~T.}\ \bibnamefont {Uysal}},\ and\ \bibinfo {author}
  {\bibfnamefont {J.~D.}\ \bibnamefont {Thompson}},\ }\bibfield  {title}
  {\bibinfo {title} {Hybrid microwave-optical scanning probe for addressing
  solid-state spins in nanophotonic cavities},\ }\href
  {https://doi.org/10.1364/OE.417528} {\bibfield  {journal} {\bibinfo
  {journal} {Opt. Express}\ }\textbf {\bibinfo {volume} {29}},\ \bibinfo
  {pages} {4902} (\bibinfo {year} {2021})}\BibitemShut {NoStop}%
\bibitem [{\citenamefont {Kolkowitz}\ \emph {et~al.}(2012)\citenamefont
  {Kolkowitz}, \citenamefont {Unterreithmeier}, \citenamefont {Bennett},\ and\
  \citenamefont {Lukin}}]{kolkowitz2012sensing}%
  \BibitemOpen
  \bibfield  {author} {\bibinfo {author} {\bibfnamefont {S.}~\bibnamefont
  {Kolkowitz}}, \bibinfo {author} {\bibfnamefont {Q.~P.}\ \bibnamefont
  {Unterreithmeier}}, \bibinfo {author} {\bibfnamefont {S.~D.}\ \bibnamefont
  {Bennett}},\ and\ \bibinfo {author} {\bibfnamefont {M.~D.}\ \bibnamefont
  {Lukin}},\ }\bibfield  {title} {\bibinfo {title} {Sensing distant nuclear
  spins with a single electron spin},\ }\href
  {https://doi.org/10.1103/PhysRevLett.109.137601} {\bibfield  {journal}
  {\bibinfo  {journal} {Phys. Rev. Lett.}\ }\textbf {\bibinfo {volume} {109}},\
  \bibinfo {pages} {137601} (\bibinfo {year} {2012})}\BibitemShut {NoStop}%
\bibitem [{\citenamefont {Taminiau}\ \emph {et~al.}(2012)\citenamefont
  {Taminiau}, \citenamefont {Wagenaar}, \citenamefont {Van~der Sar},
  \citenamefont {Jelezko}, \citenamefont {Dobrovitski},\ and\ \citenamefont
  {Hanson}}]{taminiau2012detection}%
  \BibitemOpen
  \bibfield  {author} {\bibinfo {author} {\bibfnamefont {T.}~\bibnamefont
  {Taminiau}}, \bibinfo {author} {\bibfnamefont {J.}~\bibnamefont {Wagenaar}},
  \bibinfo {author} {\bibfnamefont {T.}~\bibnamefont {Van~der Sar}}, \bibinfo
  {author} {\bibfnamefont {F.}~\bibnamefont {Jelezko}}, \bibinfo {author}
  {\bibfnamefont {V.~V.}\ \bibnamefont {Dobrovitski}},\ and\ \bibinfo {author}
  {\bibfnamefont {R.}~\bibnamefont {Hanson}},\ }\bibfield  {title} {\bibinfo
  {title} {Detection and control of individual nuclear spins using a weakly
  coupled electron spin},\ }\href
  {https://doi.org/10.1103/PhysRevLett.109.137602} {\bibfield  {journal}
  {\bibinfo  {journal} {Phys. Rev. Lett.}\ }\textbf {\bibinfo {volume} {109}},\
  \bibinfo {pages} {137602} (\bibinfo {year} {2012})}\BibitemShut {NoStop}%
\bibitem [{\citenamefont {Zhao}\ \emph {et~al.}(2012)\citenamefont {Zhao},
  \citenamefont {Honert}, \citenamefont {Schmid}, \citenamefont {Klas},
  \citenamefont {Isoya}, \citenamefont {Markham}, \citenamefont {Twitchen},
  \citenamefont {Jelezko}, \citenamefont {Liu}, \citenamefont {Fedder} \emph
  {et~al.}}]{zhao2012sensing}%
  \BibitemOpen
  \bibfield  {author} {\bibinfo {author} {\bibfnamefont {N.}~\bibnamefont
  {Zhao}}, \bibinfo {author} {\bibfnamefont {J.}~\bibnamefont {Honert}},
  \bibinfo {author} {\bibfnamefont {B.}~\bibnamefont {Schmid}}, \bibinfo
  {author} {\bibfnamefont {M.}~\bibnamefont {Klas}}, \bibinfo {author}
  {\bibfnamefont {J.}~\bibnamefont {Isoya}}, \bibinfo {author} {\bibfnamefont
  {M.}~\bibnamefont {Markham}}, \bibinfo {author} {\bibfnamefont
  {D.}~\bibnamefont {Twitchen}}, \bibinfo {author} {\bibfnamefont
  {F.}~\bibnamefont {Jelezko}}, \bibinfo {author} {\bibfnamefont {R.-B.}\
  \bibnamefont {Liu}}, \bibinfo {author} {\bibfnamefont {H.}~\bibnamefont
  {Fedder}}, \emph {et~al.},\ }\bibfield  {title} {\bibinfo {title} {Sensing
  single remote nuclear spins},\ }\href
  {https://doi.org/10.1038/nnano.2012.152} {\bibfield  {journal} {\bibinfo
  {journal} {Nat. Nanotechnol.}\ }\textbf {\bibinfo {volume} {7}},\ \bibinfo
  {pages} {657} (\bibinfo {year} {2012})}\BibitemShut {NoStop}%
\bibitem [{SI()}]{SI}%
  \BibitemOpen
  \href@noop {} {}\bibinfo {note} {See Supplemental Material, which contains
  Refs.~\cite{pang2005Cgrowth,kresse1996vasp1,kresse1996vasp2,heyd2003hse,oBryan1988yso_lattice},
  device parameters, theoretical background on measuring hyperfine interaction
  parameters, modeling errors during SWAP and nuclear spin position
  determination, as well as additional experimental data and theoretical
  analysis exploring shifts in the nuclear spin precession
  frequency.}\BibitemShut {Stop}%
\bibitem [{\citenamefont {Van~de Walle}\ and\ \citenamefont
  {Neugebauer}(2003)}]{vandewalle2003H}%
  \BibitemOpen
  \bibfield  {author} {\bibinfo {author} {\bibfnamefont {C.~G.}\ \bibnamefont
  {Van~de Walle}}\ and\ \bibinfo {author} {\bibfnamefont {J.}~\bibnamefont
  {Neugebauer}},\ }\bibfield  {title} {\bibinfo {title} {Universal alignment of
  hydrogen levels in semiconductors, insulators and solutions},\ }\href
  {https://doi.org/10.1038/nature01665} {\bibfield  {journal} {\bibinfo
  {journal} {Nature}\ }\textbf {\bibinfo {volume} {423}},\ \bibinfo {pages}
  {626} (\bibinfo {year} {2003})}\BibitemShut {NoStop}%
\bibitem [{\citenamefont {McCluskey}\ \emph {et~al.}(2012)\citenamefont
  {McCluskey}, \citenamefont {Tarun},\ and\ \citenamefont
  {Teklemichael}}]{mccluskey2012hydrogen}%
  \BibitemOpen
  \bibfield  {author} {\bibinfo {author} {\bibfnamefont {M.~D.}\ \bibnamefont
  {McCluskey}}, \bibinfo {author} {\bibfnamefont {M.~C.}\ \bibnamefont
  {Tarun}},\ and\ \bibinfo {author} {\bibfnamefont {S.~T.}\ \bibnamefont
  {Teklemichael}},\ }\bibfield  {title} {\bibinfo {title} {Hydrogen in oxide
  semiconductors},\ }\href {https://doi.org/10.1557/jmr.2012.137} {\bibfield
  {journal} {\bibinfo  {journal} {J. Mater. Res.}\ }\textbf {\bibinfo {volume}
  {27}},\ \bibinfo {pages} {2190} (\bibinfo {year} {2012})}\BibitemShut
  {NoStop}%
\bibitem [{\citenamefont {Shi}\ \emph {et~al.}(2004)\citenamefont {Shi},
  \citenamefont {Saboktakin}, \citenamefont {Stavola},\ and\ \citenamefont
  {Pearton}}]{shi2004hidden}%
  \BibitemOpen
  \bibfield  {author} {\bibinfo {author} {\bibfnamefont {G.~A.}\ \bibnamefont
  {Shi}}, \bibinfo {author} {\bibfnamefont {M.}~\bibnamefont {Saboktakin}},
  \bibinfo {author} {\bibfnamefont {M.}~\bibnamefont {Stavola}},\ and\ \bibinfo
  {author} {\bibfnamefont {S.}~\bibnamefont {Pearton}},\ }\bibfield  {title}
  {\bibinfo {title} {``\text{H}idden hydrogen'' in as-grown $\text{ZnO}$},\
  }\href {https://doi.org/10.1063/1.1832736} {\bibfield  {journal} {\bibinfo
  {journal} {Appl. Phys. Lett.}\ }\textbf {\bibinfo {volume} {85}},\ \bibinfo
  {pages} {5601} (\bibinfo {year} {2004})}\BibitemShut {NoStop}%
\bibitem [{\citenamefont {Devor}\ \emph {et~al.}(1984)\citenamefont {Devor},
  \citenamefont {Pastor},\ and\ \citenamefont {DeShazer}}]{devor1984hydroxyl}%
  \BibitemOpen
  \bibfield  {author} {\bibinfo {author} {\bibfnamefont {D.}~\bibnamefont
  {Devor}}, \bibinfo {author} {\bibfnamefont {R.}~\bibnamefont {Pastor}},\ and\
  \bibinfo {author} {\bibfnamefont {L.}~\bibnamefont {DeShazer}},\ }\bibfield
  {title} {\bibinfo {title} {Hydroxyl impurity effects in $\text{YAG}$
  ($\text{Y}_3\text{Al}_5\text{O}_{12}$)},\ }\href
  {https://doi.org/10.1063/1.448156} {\bibfield  {journal} {\bibinfo  {journal}
  {J. Chem. Phys.}\ }\textbf {\bibinfo {volume} {81}},\ \bibinfo {pages} {4104}
  (\bibinfo {year} {1984})}\BibitemShut {NoStop}%
\bibitem [{\citenamefont {Mu}\ \emph {et~al.}(2022)\citenamefont {Mu},
  \citenamefont {Wang}, \citenamefont {Varley}, \citenamefont {Lyons},
  \citenamefont {Wickramaratne},\ and\ \citenamefont {Van~de
  Walle}}]{mu2022role}%
  \BibitemOpen
  \bibfield  {author} {\bibinfo {author} {\bibfnamefont {S.}~\bibnamefont
  {Mu}}, \bibinfo {author} {\bibfnamefont {M.}~\bibnamefont {Wang}}, \bibinfo
  {author} {\bibfnamefont {J.~B.}\ \bibnamefont {Varley}}, \bibinfo {author}
  {\bibfnamefont {J.~L.}\ \bibnamefont {Lyons}}, \bibinfo {author}
  {\bibfnamefont {D.}~\bibnamefont {Wickramaratne}},\ and\ \bibinfo {author}
  {\bibfnamefont {C.~G.}\ \bibnamefont {Van~de Walle}},\ }\bibfield  {title}
  {\bibinfo {title} {Role of carbon and hydrogen in limiting $n$-type doping of
  monoclinic $(\text{Al}_x\text{Ga}_{1-x})_2\text{O}_3$},\ }\href
  {https://journals.aps.org/prb/abstract/10.1103/PhysRevB.105.155201}
  {\bibfield  {journal} {\bibinfo  {journal} {Phys. Rev. B}\ }\textbf {\bibinfo
  {volume} {105}},\ \bibinfo {pages} {155201} (\bibinfo {year}
  {2022})}\BibitemShut {NoStop}%
\bibitem [{\citenamefont {Van~der Sar}\ \emph {et~al.}(2012)\citenamefont
  {Van~der Sar}, \citenamefont {Wang}, \citenamefont {Blok}, \citenamefont
  {Bernien}, \citenamefont {Taminiau}, \citenamefont {Toyli}, \citenamefont
  {Lidar}, \citenamefont {Awschalom}, \citenamefont {Hanson},\ and\
  \citenamefont {Dobrovitski}}]{van2012decoherence}%
  \BibitemOpen
  \bibfield  {author} {\bibinfo {author} {\bibfnamefont {T.}~\bibnamefont
  {Van~der Sar}}, \bibinfo {author} {\bibfnamefont {Z.}~\bibnamefont {Wang}},
  \bibinfo {author} {\bibfnamefont {M.}~\bibnamefont {Blok}}, \bibinfo {author}
  {\bibfnamefont {H.}~\bibnamefont {Bernien}}, \bibinfo {author} {\bibfnamefont
  {T.}~\bibnamefont {Taminiau}}, \bibinfo {author} {\bibfnamefont
  {D.}~\bibnamefont {Toyli}}, \bibinfo {author} {\bibfnamefont
  {D.}~\bibnamefont {Lidar}}, \bibinfo {author} {\bibfnamefont
  {D.}~\bibnamefont {Awschalom}}, \bibinfo {author} {\bibfnamefont
  {R.}~\bibnamefont {Hanson}},\ and\ \bibinfo {author} {\bibfnamefont
  {V.}~\bibnamefont {Dobrovitski}},\ }\bibfield  {title} {\bibinfo {title}
  {Decoherence-protected quantum gates for a hybrid solid-state spin
  register},\ }\href {https://doi.org/10.1038/nature10900} {\bibfield
  {journal} {\bibinfo  {journal} {Nature}\ }\textbf {\bibinfo {volume} {484}},\
  \bibinfo {pages} {82} (\bibinfo {year} {2012})}\BibitemShut {NoStop}%
\bibitem [{\citenamefont {Bradley}\ \emph {et~al.}(2019)\citenamefont
  {Bradley}, \citenamefont {Randall}, \citenamefont {Abobeih}, \citenamefont
  {Berrevoets}, \citenamefont {Degen}, \citenamefont {Bakker}, \citenamefont
  {Markham}, \citenamefont {Twitchen},\ and\ \citenamefont
  {Taminiau}}]{bradley2019ten}%
  \BibitemOpen
  \bibfield  {author} {\bibinfo {author} {\bibfnamefont {C.~E.}\ \bibnamefont
  {Bradley}}, \bibinfo {author} {\bibfnamefont {J.}~\bibnamefont {Randall}},
  \bibinfo {author} {\bibfnamefont {M.~H.}\ \bibnamefont {Abobeih}}, \bibinfo
  {author} {\bibfnamefont {R.~C.}\ \bibnamefont {Berrevoets}}, \bibinfo
  {author} {\bibfnamefont {M.~J.}\ \bibnamefont {Degen}}, \bibinfo {author}
  {\bibfnamefont {M.~A.}\ \bibnamefont {Bakker}}, \bibinfo {author}
  {\bibfnamefont {M.}~\bibnamefont {Markham}}, \bibinfo {author} {\bibfnamefont
  {D.~J.}\ \bibnamefont {Twitchen}},\ and\ \bibinfo {author} {\bibfnamefont
  {T.~H.}\ \bibnamefont {Taminiau}},\ }\bibfield  {title} {\bibinfo {title} {A
  ten-qubit solid-state spin register with quantum memory up to one minute},\
  }\href {https://doi.org/10.1103/PhysRevX.9.031045} {\bibfield  {journal}
  {\bibinfo  {journal} {Phys. Rev. X}\ }\textbf {\bibinfo {volume} {9}},\
  \bibinfo {pages} {031045} (\bibinfo {year} {2019})}\BibitemShut {NoStop}%
\bibitem [{\citenamefont {Kalb}\ \emph {et~al.}(2018)\citenamefont {Kalb},
  \citenamefont {Humphreys}, \citenamefont {Slim},\ and\ \citenamefont
  {Hanson}}]{kalb2018dephasing}%
  \BibitemOpen
  \bibfield  {author} {\bibinfo {author} {\bibfnamefont {N.}~\bibnamefont
  {Kalb}}, \bibinfo {author} {\bibfnamefont {P.~C.}\ \bibnamefont {Humphreys}},
  \bibinfo {author} {\bibfnamefont {J.}~\bibnamefont {Slim}},\ and\ \bibinfo
  {author} {\bibfnamefont {R.}~\bibnamefont {Hanson}},\ }\bibfield  {title}
  {\bibinfo {title} {Dephasing mechanisms of diamond-based nuclear-spin
  memories for quantum networks},\ }\href
  {https://doi.org/10.1103/PhysRevA.97.062330} {\bibfield  {journal} {\bibinfo
  {journal} {Phys. Rev. A}\ }\textbf {\bibinfo {volume} {97}},\ \bibinfo
  {pages} {062330} (\bibinfo {year} {2018})}\BibitemShut {NoStop}%
\bibitem [{\citenamefont {Reiserer}\ \emph {et~al.}(2016)\citenamefont
  {Reiserer}, \citenamefont {Kalb}, \citenamefont {Blok}, \citenamefont {van
  Bemmelen}, \citenamefont {Taminiau}, \citenamefont {Hanson}, \citenamefont
  {Twitchen},\ and\ \citenamefont {Markham}}]{reiserer2016robust}%
  \BibitemOpen
  \bibfield  {author} {\bibinfo {author} {\bibfnamefont {A.}~\bibnamefont
  {Reiserer}}, \bibinfo {author} {\bibfnamefont {N.}~\bibnamefont {Kalb}},
  \bibinfo {author} {\bibfnamefont {M.~S.}\ \bibnamefont {Blok}}, \bibinfo
  {author} {\bibfnamefont {K.~J.}\ \bibnamefont {van Bemmelen}}, \bibinfo
  {author} {\bibfnamefont {T.~H.}\ \bibnamefont {Taminiau}}, \bibinfo {author}
  {\bibfnamefont {R.}~\bibnamefont {Hanson}}, \bibinfo {author} {\bibfnamefont
  {D.~J.}\ \bibnamefont {Twitchen}},\ and\ \bibinfo {author} {\bibfnamefont
  {M.}~\bibnamefont {Markham}},\ }\bibfield  {title} {\bibinfo {title} {Robust
  quantum-network memory using decoherence-protected subspaces of nuclear
  spins},\ }\href {https://doi.org/10.1103/PhysRevX.6.021040} {\bibfield
  {journal} {\bibinfo  {journal} {Phys. Rev. X}\ }\textbf {\bibinfo {volume}
  {6}},\ \bibinfo {pages} {021040} (\bibinfo {year} {2016})}\BibitemShut
  {NoStop}%
\bibitem [{\citenamefont {Pang}\ \emph {et~al.}(2005)\citenamefont {Pang},
  \citenamefont {Zhao}, \citenamefont {Jie}, \citenamefont {Xu},\ and\
  \citenamefont {He}}]{pang2005Cgrowth}%
  \BibitemOpen
  \bibfield  {author} {\bibinfo {author} {\bibfnamefont {H.}~\bibnamefont
  {Pang}}, \bibinfo {author} {\bibfnamefont {G.}~\bibnamefont {Zhao}}, \bibinfo
  {author} {\bibfnamefont {M.}~\bibnamefont {Jie}}, \bibinfo {author}
  {\bibfnamefont {J.}~\bibnamefont {Xu}},\ and\ \bibinfo {author}
  {\bibfnamefont {X.}~\bibnamefont {He}},\ }\bibfield  {title} {\bibinfo
  {title} {Study on the growth, etch morphology and spectra of
  $\text{Y}_2\text{SiO}_5$ crystal},\ }\href
  {https://doi.org/10.1016/j.matlet.2005.06.036} {\bibfield  {journal}
  {\bibinfo  {journal} {Mater. Lett.}\ }\textbf {\bibinfo {volume} {59}},\
  \bibinfo {pages} {3539} (\bibinfo {year} {2005})}\BibitemShut {NoStop}%
\bibitem [{\citenamefont {Kresse}\ and\ \citenamefont
  {Furthm{\"u}ller}(1996{\natexlab{a}})}]{kresse1996vasp1}%
  \BibitemOpen
  \bibfield  {author} {\bibinfo {author} {\bibfnamefont {G.}~\bibnamefont
  {Kresse}}\ and\ \bibinfo {author} {\bibfnamefont {J.}~\bibnamefont
  {Furthm{\"u}ller}},\ }\bibfield  {title} {\bibinfo {title} {Efficient
  iterative schemes for ab initio total-energy calculations using a plane-wave
  basis set},\ }\href {https://doi.org/10.1103/PhysRevB.54.11169} {\bibfield
  {journal} {\bibinfo  {journal} {Phys. Rev. B}\ }\textbf {\bibinfo {volume}
  {54}},\ \bibinfo {pages} {11169} (\bibinfo {year}
  {1996}{\natexlab{a}})}\BibitemShut {NoStop}%
\bibitem [{\citenamefont {Kresse}\ and\ \citenamefont
  {Furthm{\"u}ller}(1996{\natexlab{b}})}]{kresse1996vasp2}%
  \BibitemOpen
  \bibfield  {author} {\bibinfo {author} {\bibfnamefont {G.}~\bibnamefont
  {Kresse}}\ and\ \bibinfo {author} {\bibfnamefont {J.}~\bibnamefont
  {Furthm{\"u}ller}},\ }\bibfield  {title} {\bibinfo {title} {Efficiency of
  ab-initio total energy calculations for metals and semiconductors using a
  plane-wave basis set},\ }\href {https://doi.org/10.1016/0927-0256(96)00008-0}
  {\bibfield  {journal} {\bibinfo  {journal} {Comput. Mater. Sci.}\ }\textbf
  {\bibinfo {volume} {6}},\ \bibinfo {pages} {15} (\bibinfo {year}
  {1996}{\natexlab{b}})}\BibitemShut {NoStop}%
\bibitem [{\citenamefont {Heyd}\ \emph {et~al.}(2003)\citenamefont {Heyd},
  \citenamefont {Scuseria},\ and\ \citenamefont {Ernzerhof}}]{heyd2003hse}%
  \BibitemOpen
  \bibfield  {author} {\bibinfo {author} {\bibfnamefont {J.}~\bibnamefont
  {Heyd}}, \bibinfo {author} {\bibfnamefont {G.~E.}\ \bibnamefont {Scuseria}},\
  and\ \bibinfo {author} {\bibfnamefont {M.}~\bibnamefont {Ernzerhof}},\
  }\bibfield  {title} {\bibinfo {title} {Hybrid functionals based on a screened
  $\text{C}$oulomb potential},\ }\href {https://doi.org/10.1063/1.1564060}
  {\bibfield  {journal} {\bibinfo  {journal} {J. Chem. Phys.}\ }\textbf
  {\bibinfo {volume} {118}},\ \bibinfo {pages} {8207} (\bibinfo {year}
  {2003})}\BibitemShut {NoStop}%
\bibitem [{\citenamefont {O'Bryan}\ \emph {et~al.}(1988)\citenamefont
  {O'Bryan}, \citenamefont {Gallagher},\ and\ \citenamefont
  {Berkstresser}}]{oBryan1988yso_lattice}%
  \BibitemOpen
  \bibfield  {author} {\bibinfo {author} {\bibfnamefont {H.~M.}\ \bibnamefont
  {O'Bryan}}, \bibinfo {author} {\bibfnamefont {P.~K.}\ \bibnamefont
  {Gallagher}},\ and\ \bibinfo {author} {\bibfnamefont {G.}~\bibnamefont
  {Berkstresser}},\ }\bibfield  {title} {\bibinfo {title} {Thermal expansion of
  $\text{Y}_2\text{SiO}_5$ single crystals},\ }\href
  {https://doi.org/10.1111/j.1151-2916.1988.tb05779.x} {\bibfield  {journal}
  {\bibinfo  {journal} {J. Am. Ceram. Soc.}\ }\textbf {\bibinfo {volume}
  {71}},\ \bibinfo {pages} {C42} (\bibinfo {year} {1988})}\BibitemShut
  {NoStop}%
\end{thebibliography}%


\begin{thebibliography}{9}%
\makeatletter
\providecommand \@ifxundefined [1]{%
 \@ifx{#1\undefined}
}%
\providecommand \@ifnum [1]{%
 \ifnum #1\expandafter \@firstoftwo
 \else \expandafter \@secondoftwo
 \fi
}%
\providecommand \@ifx [1]{%
 \ifx #1\expandafter \@firstoftwo
 \else \expandafter \@secondoftwo
 \fi
}%
\providecommand \natexlab [1]{#1}%
\providecommand \enquote  [1]{``#1''}%
\providecommand \bibnamefont  [1]{#1}%
\providecommand \bibfnamefont [1]{#1}%
\providecommand \citenamefont [1]{#1}%
\providecommand \href@noop [0]{\@secondoftwo}%
\providecommand \href [0]{\begingroup \@sanitize@url \@href}%
\providecommand \@href[1]{\@@startlink{#1}\@@href}%
\providecommand \@@href[1]{\endgroup#1\@@endlink}%
\providecommand \@sanitize@url [0]{\catcode `\\12\catcode `\$12\catcode
  `\&12\catcode `\#12\catcode `\^12\catcode `\_12\catcode `\%12\relax}%
\providecommand \@@startlink[1]{}%
\providecommand \@@endlink[0]{}%
\providecommand \url  [0]{\begingroup\@sanitize@url \@url }%
\providecommand \@url [1]{\endgroup\@href {#1}{\urlprefix }}%
\providecommand \urlprefix  [0]{URL }%
\providecommand \Eprint [0]{\href }%
\providecommand \doibase [0]{https://doi.org/}%
\providecommand \selectlanguage [0]{\@gobble}%
\providecommand \bibinfo  [0]{\@secondoftwo}%
\providecommand \bibfield  [0]{\@secondoftwo}%
\providecommand \translation [1]{[#1]}%
\providecommand \BibitemOpen [0]{}%
\providecommand \bibitemStop [0]{}%
\providecommand \bibitemNoStop [0]{.\EOS\space}%
\providecommand \EOS [0]{\spacefactor3000\relax}%
\providecommand \BibitemShut  [1]{\csname bibitem#1\endcsname}%
\let\auto@bib@innerbib\@empty
\bibitem [{\citenamefont {Chen}\ \emph {et~al.}(2020)\citenamefont {Chen},
  \citenamefont {Raha}, \citenamefont {Phenicie}, \citenamefont {Ourari},\ and\
  \citenamefont {Thompson}}]{chen2020parallel}%
  \BibitemOpen
  \bibfield  {author} {\bibinfo {author} {\bibfnamefont {S.}~\bibnamefont
  {Chen}}, \bibinfo {author} {\bibfnamefont {M.}~\bibnamefont {Raha}}, \bibinfo
  {author} {\bibfnamefont {C.~M.}\ \bibnamefont {Phenicie}}, \bibinfo {author}
  {\bibfnamefont {S.}~\bibnamefont {Ourari}},\ and\ \bibinfo {author}
  {\bibfnamefont {J.~D.}\ \bibnamefont {Thompson}},\ }\href
  {https://doi.org/10.1126/science.abc7821} {\bibfield  {journal} {\bibinfo
  {journal} {Science}\ }\textbf {\bibinfo {volume} {370}},\ \bibinfo {pages}
  {592} (\bibinfo {year} {2020})}\BibitemShut {NoStop}%
\bibitem [{\citenamefont {Chen}\ \emph {et~al.}(2021)\citenamefont {Chen},
  \citenamefont {Ourari}, \citenamefont {Raha}, \citenamefont {Phenicie},
  \citenamefont {Uysal},\ and\ \citenamefont {Thompson}}]{chen2021hybrid}%
  \BibitemOpen
  \bibfield  {author} {\bibinfo {author} {\bibfnamefont {S.}~\bibnamefont
  {Chen}}, \bibinfo {author} {\bibfnamefont {S.}~\bibnamefont {Ourari}},
  \bibinfo {author} {\bibfnamefont {M.}~\bibnamefont {Raha}}, \bibinfo {author}
  {\bibfnamefont {C.~M.}\ \bibnamefont {Phenicie}}, \bibinfo {author}
  {\bibfnamefont {M.~T.}\ \bibnamefont {Uysal}},\ and\ \bibinfo {author}
  {\bibfnamefont {J.~D.}\ \bibnamefont {Thompson}},\ }\href
  {https://doi.org/10.1364/OE.417528} {\bibfield  {journal} {\bibinfo
  {journal} {Opt. Express}\ }\textbf {\bibinfo {volume} {29}},\ \bibinfo
  {pages} {4902} (\bibinfo {year} {2021})}\BibitemShut {NoStop}%
\bibitem [{\citenamefont {Sun}\ \emph {et~al.}(2008)\citenamefont {Sun},
  \citenamefont {B{\"o}ttger}, \citenamefont {Thiel},\ and\ \citenamefont
  {Cone}}]{sun2008magnetic}%
  \BibitemOpen
  \bibfield  {author} {\bibinfo {author} {\bibfnamefont {Y.}~\bibnamefont
  {Sun}}, \bibinfo {author} {\bibfnamefont {T.}~\bibnamefont {B{\"o}ttger}},
  \bibinfo {author} {\bibfnamefont {C.}~\bibnamefont {Thiel}},\ and\ \bibinfo
  {author} {\bibfnamefont {R.}~\bibnamefont {Cone}},\ }\href
  {https://doi.org/10.1103/PhysRevB.77.085124} {\bibfield  {journal} {\bibinfo
  {journal} {Phys. Rev. B}\ }\textbf {\bibinfo {volume} {77}},\ \bibinfo
  {pages} {085124} (\bibinfo {year} {2008})}\BibitemShut {NoStop}%
\bibitem [{\citenamefont {Degen}\ \emph {et~al.}(2021)\citenamefont {Degen},
  \citenamefont {Loenen}, \citenamefont {Bartling}, \citenamefont {Bradley},
  \citenamefont {Meinsma}, \citenamefont {Markham}, \citenamefont {Twitchen},\
  and\ \citenamefont {Taminiau}}]{degen2021entanglement}%
  \BibitemOpen
  \bibfield  {author} {\bibinfo {author} {\bibfnamefont {M.}~\bibnamefont
  {Degen}}, \bibinfo {author} {\bibfnamefont {S.}~\bibnamefont {Loenen}},
  \bibinfo {author} {\bibfnamefont {H.}~\bibnamefont {Bartling}}, \bibinfo
  {author} {\bibfnamefont {C.}~\bibnamefont {Bradley}}, \bibinfo {author}
  {\bibfnamefont {A.}~\bibnamefont {Meinsma}}, \bibinfo {author} {\bibfnamefont
  {M.}~\bibnamefont {Markham}}, \bibinfo {author} {\bibfnamefont
  {D.}~\bibnamefont {Twitchen}},\ and\ \bibinfo {author} {\bibfnamefont
  {T.}~\bibnamefont {Taminiau}},\ }\href
  {https://doi.org/10.1038/s41467-021-23454-9} {\bibfield  {journal} {\bibinfo
  {journal} {Nat. Commun.}\ }\textbf {\bibinfo {volume} {12}},\ \bibinfo
  {pages} {3470} (\bibinfo {year} {2021})}\BibitemShut {NoStop}%
\bibitem [{\citenamefont {Kresse}\ and\ \citenamefont
  {Furthm{\"u}ller}(1996{\natexlab{a}})}]{kresse1996vasp1}%
  \BibitemOpen
  \bibfield  {author} {\bibinfo {author} {\bibfnamefont {G.}~\bibnamefont
  {Kresse}}\ and\ \bibinfo {author} {\bibfnamefont {J.}~\bibnamefont
  {Furthm{\"u}ller}},\ }\href {https://doi.org/10.1103/PhysRevB.54.11169}
  {\bibfield  {journal} {\bibinfo  {journal} {Phys. Rev. B}\ }\textbf {\bibinfo
  {volume} {54}},\ \bibinfo {pages} {11169} (\bibinfo {year}
  {1996}{\natexlab{a}})}\BibitemShut {NoStop}%
\bibitem [{\citenamefont {Kresse}\ and\ \citenamefont
  {Furthm{\"u}ller}(1996{\natexlab{b}})}]{kresse1996vasp2}%
  \BibitemOpen
  \bibfield  {author} {\bibinfo {author} {\bibfnamefont {G.}~\bibnamefont
  {Kresse}}\ and\ \bibinfo {author} {\bibfnamefont {J.}~\bibnamefont
  {Furthm{\"u}ller}},\ }\href {https://doi.org/10.1016/0927-0256(96)00008-0}
  {\bibfield  {journal} {\bibinfo  {journal} {Comput. Mater. Sci.}\ }\textbf
  {\bibinfo {volume} {6}},\ \bibinfo {pages} {15} (\bibinfo {year}
  {1996}{\natexlab{b}})}\BibitemShut {NoStop}%
\bibitem [{\citenamefont {Heyd}\ \emph {et~al.}(2003)\citenamefont {Heyd},
  \citenamefont {Scuseria},\ and\ \citenamefont {Ernzerhof}}]{heyd2003hse}%
  \BibitemOpen
  \bibfield  {author} {\bibinfo {author} {\bibfnamefont {J.}~\bibnamefont
  {Heyd}}, \bibinfo {author} {\bibfnamefont {G.~E.}\ \bibnamefont {Scuseria}},\
  and\ \bibinfo {author} {\bibfnamefont {M.}~\bibnamefont {Ernzerhof}},\ }\href
  {https://aip.scitation.org/doi/abs/10.1063/1.1564060} {\bibfield  {journal}
  {\bibinfo  {journal} {J. Chem. Phys.}\ }\textbf {\bibinfo {volume} {118}},\
  \bibinfo {pages} {8207} (\bibinfo {year} {2003})}\BibitemShut {NoStop}%
\bibitem [{\citenamefont {Pang}\ \emph {et~al.}(2005)\citenamefont {Pang},
  \citenamefont {Zhao}, \citenamefont {Jie}, \citenamefont {Xu},\ and\
  \citenamefont {He}}]{pang2005Cgrowth}%
  \BibitemOpen
  \bibfield  {author} {\bibinfo {author} {\bibfnamefont {H.}~\bibnamefont
  {Pang}}, \bibinfo {author} {\bibfnamefont {G.}~\bibnamefont {Zhao}}, \bibinfo
  {author} {\bibfnamefont {M.}~\bibnamefont {Jie}}, \bibinfo {author}
  {\bibfnamefont {J.}~\bibnamefont {Xu}},\ and\ \bibinfo {author}
  {\bibfnamefont {X.}~\bibnamefont {He}},\ }\href
  {https://www.sciencedirect.com/science/article/abs/pii/S0167577X05006191}
  {\bibfield  {journal} {\bibinfo  {journal} {Mater. Lett.}\ }\textbf {\bibinfo
  {volume} {59}},\ \bibinfo {pages} {3539} (\bibinfo {year}
  {2005})}\BibitemShut {NoStop}%
\bibitem [{\citenamefont {O'Bryan}\ \emph {et~al.}(1988)\citenamefont
  {O'Bryan}, \citenamefont {Gallagher},\ and\ \citenamefont
  {Berkstresser}}]{oBryan1988yso_lattice}%
  \BibitemOpen
  \bibfield  {author} {\bibinfo {author} {\bibfnamefont {H.~M.}\ \bibnamefont
  {O'Bryan}}, \bibinfo {author} {\bibfnamefont {P.~K.}\ \bibnamefont
  {Gallagher}},\ and\ \bibinfo {author} {\bibfnamefont {G.}~\bibnamefont
  {Berkstresser}},\ }\href
  {https://ceramics.onlinelibrary.wiley.com/doi/10.1111/j.1151-2916.1988.tb05779.x}
  {\bibfield  {journal} {\bibinfo  {journal} {J. Am. Ceram. Soc.}\ }\textbf
  {\bibinfo {volume} {71}},\ \bibinfo {pages} {C42} (\bibinfo {year}
  {1988})}\BibitemShut {NoStop}%
\end{thebibliography}%


%

\end{document}